\documentclass[twocolumn,prb,preprintnumbers,showpacs,floatfix,superscriptaddress]{revtex4}

\usepackage{amssymb}
\usepackage{graphicx}
\usepackage{dcolumn}
\usepackage{bm}
\usepackage{sidecap}

\begin{document}

\title{Multiplet ligand-field theory using Wannier orbitals.}
\author{M. W. Haverkort}
\affiliation{Max Planck Institute for Solid State Research, Heisenbergstra{\ss }e 1,
70569 Stuttgart, Germany}
\author{M. Zwierzycki}
\affiliation{Institute of Molecular Physics, Polish Academy of Sciences, M.
Smoluchowskiego 17, 60-179 Pozna\'n, Poland}
\author{O. K. Andersen}
\affiliation{Max Planck Institute for Solid State Research, Heisenbergstra{\ss }e 1,
70569 Stuttgart, Germany}
\date{\today}

\begin{abstract}
We demonstrate how \textit{ab initio} cluster calculations including the
full Coulomb vertex can be done in the basis of the localized
Wannier orbitals which describe the low-energy density functional (LDA) band
structure of the infinite crystal, \emph{e.g.} the transition metal $3d$ and
oxygen $2p$ orbitals. The spatial extend of our $3d$ Wannier orbitals
(orthonormalized $N$th order muffin-tin orbitals) is close to that found for
atomic Hartree-Fock orbitals. We define Ligand orbitals as
those linear combinations of the O $2p$ Wannier orbitals which couple to the 
$3d$ orbitals for the chosen cluster. The use of ligand orbitals allows for
a minimal Hilbert space in multiplet ligand-field theory calculations, thus
reducing the computational costs substantially. The result is a fast and
simple \textit{ab initio} theory, which can provide useful information about
local properties of correlated insulators. We compare results for NiO, MnO
and SrTiO$_{3}$ with x-ray absorption, inelastic x-ray scattering, and
photoemission experiments. The multiplet ligand field theory parameters
found by our \textit{ab initio} method agree within $\sim $10\% to known
experimental values.
\end{abstract}

\maketitle


\begin{SCfigure}[][b]
    \label{NiO6Cluster}
    \includegraphics[width=0.2\textwidth]{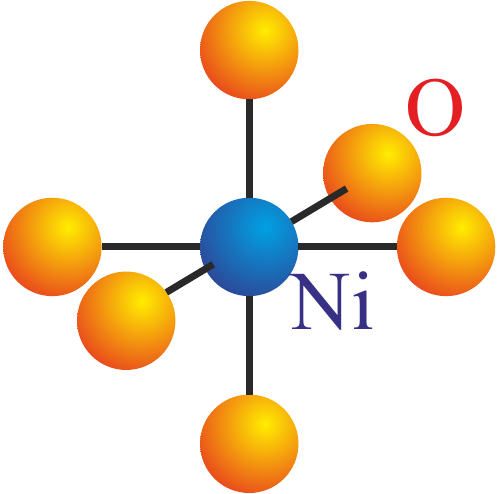}
    \caption{(color online) A NiO$_{6}$ cluster used in Multiplet Ligand Field Theory (MLFT) as a local representation of the rock salt face-centered cubic NiO solid. The Ni cation is surrounded by its 6 nearest O ligands.}
    \label{NiO6Cluster}
 \end{SCfigure}

\begin{figure*}[t]
\includegraphics[width=0.95\textwidth]{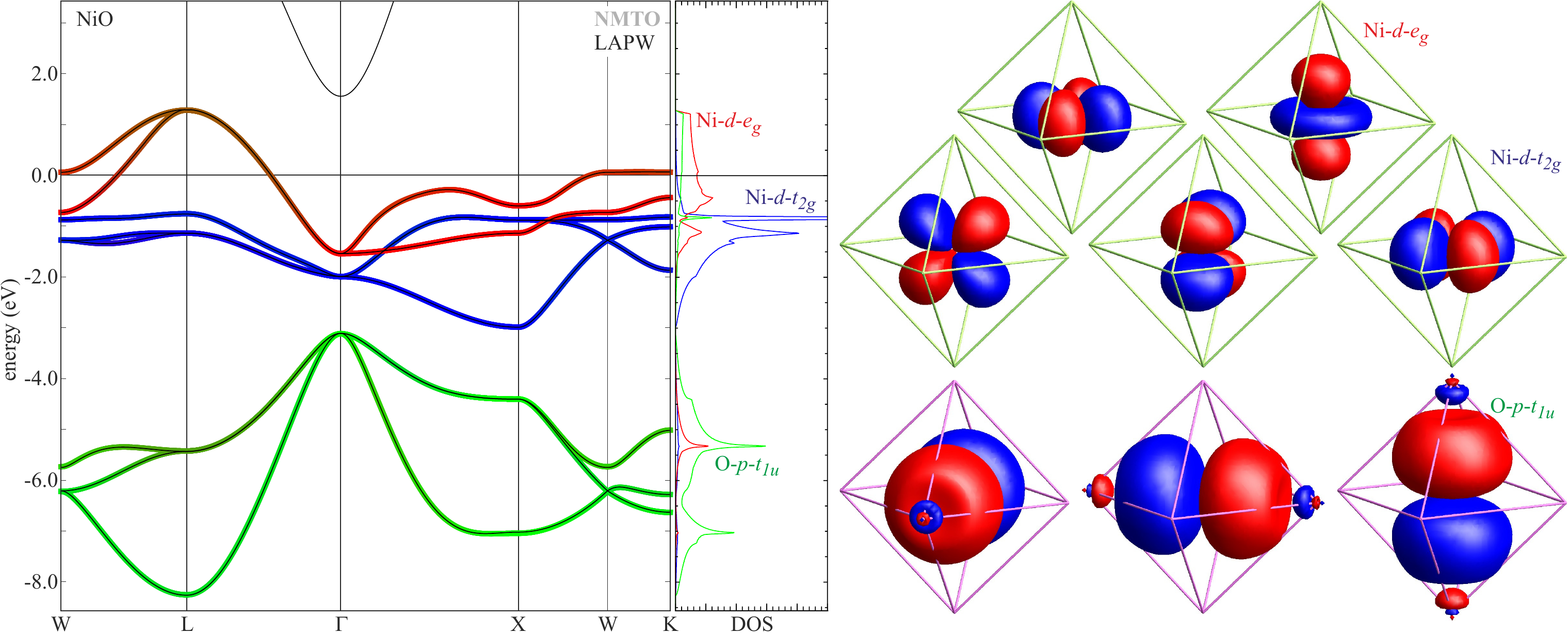}
\caption{(color online) Left panel: NiO LDA band structure calculated with
the large LAPW basis set (thin, black lines) and with the small
Wannier-orbital basis set consisting of 3 O $p$ (green), 3 Ni $d\left(
t_{2g}\right) $ (blue), and 2 Ni $d\left( e_{g}\right) $ (red)
orthonormalized $N$MTOs per NiO (thick colored lines). Colors are mixed
according to the hybridization between the Bloch sums of the three kinds of
orbitals. The Fermi level is taken as the zero of energy. Middle panel:
Wannier-orbital projected densities of states. Right panel: The eight
Wannier orbitals. Shown are constant-amplitude contours containing 90\% of
the orbital charge with the color (red or blue) giving the sign. The
Ni(O)-centered octahedra have O(Ni) at their corners.}
\label{NiOLDA}
\end{figure*}

Many electronic properties of solids can now be described \textit{ab initio}
thanks to the advent of powerful computers and the development of ingenious
methods, such as density-functional theory (DFT)\cite{Hohenberg64, Kohn65,
Kohn99} with local density (LDA)\cite{Ceperley80} or generalized gradient
(GGA)\cite{Perdew96} approximations, LDA+Hubbard $U$ (LDA+U),\cite{Anisimov91A, Anisimov93} quantum chemical methods,\cite{Graaf96, Bande08, Hozoi09, Fulde02, Neese07} dynamic mean-field theory,\cite%
{Metzner89,Georges92,Jarrell92,Georges96,Anisimov97, Held03, Maier05,
Lechermann06} quantum Monte-Carlo simulations,\cite{Foulkes01,Dagotto94} and
exact diagonalization for finite clusters.\cite{Dagotto94, Weisse06}
Nevertheless, for correlated open-shell systems with several local orbital
and spin degrees of freedom, electronic-structure calculations remain a
challenge.

Ground-state properties and spectral functions may be calculated by exact
diagonalization of the many-electron Hamiltonian, but this is hindered by
the exponential growth of the Hilbert-space with the number of correlated
electrons in the system. Exploiting symmetry and limiting the number of
correlated electronic degrees of freedom may enable the treatment of
relatively large clusters, as done in the important case of doped high-$%
T_{c} $ cuprates, where symmetry in the spin sector allowed Lau and coworkers to use
clusters with up to 32 CuO$_{2}$ plackets, each with a single Cu 
$d_{x^{2}-y^{2}}$, and two O $p$ orbitals.\cite{Lau11} For \emph{local}
properties, such as excitonic spectra, exact diagonalization for finite
clusters becomes much more appealing, as relatively small
clusters often suffice. Magnetic anisotropies, $g$-tensors,
magnetization-dependent electron-spin resonance spectra, crystal-field
excitations, and a manifold of excitonic core-level spectra are usually well
described using very small clusters. For transition-metal and rare-earth
compounds, the cluster may often be limited to merely a single $d$- or $f$%
-electron cation surrounded by its nearest neighbor ligands as illustrated
in Fig.$\,$\ref{NiO6Cluster}. For clusters that small, exact diagonalization
is equivalent to multiplet ligand-field-theory (MLFT), one of the earliest
quantum-chemistry methods developed to describe the electronic structure of
transition-metal and rare-earth compounds.\cite{Ballhausen62} MLFT is a
highly cost-efficient method, able to account for many of the local
properties and excitonic spectra of correlated materials.

MLFT calculations traditionally use parameters fitted to experiments.
Despite being a great help for understanding and interpreting experimental
results, this approach is however not completely satisfactory and, over the
years, numerous theoretical studies have therefore been devoted to obtaining
MLFT parameters \textit{ab initio}.\cite{Sugano63, Wachters71, Wachters72,
Mulliken66, Pople99, Brener87, Ogasawara01, Ikeno05, Ikeno06, Ikeno09,
Miedema11, Ikeno11, Graaf98, Geleijns99, Sadoc07,Hozoi10} Sugano and Shulman 
\cite{Sugano63} calculated the ligand-field parameters by constructing
single-particle molecular orbitals (MOs) as linear combinations of atomic
Hartree-Fock orbitals and thereby in several cases obtained qualitative
agreement with experiments. More often, MO theory with a more complete basis
is used.\cite{Mulliken66, Pople99} After the LDA had proven useful not only
for $s$- and $p$-, but also for $d$- and $f$- electrons in solids,\cite%
{Andersen70} several authors obtained MLFT parameters by performing an LDA
calculation for the cluster and using its Kohn-Sham MOs. \cite{Brener87,
Ogasawara01, Ikeno05, Ikeno06, Ikeno09, Miedema11, Ikeno11} Such a
calculation breaks the translation invariance of the crystal already at the
single-particle LDA level, and it is necessary to remedy finite-size and
surface effects, e.g. by embedding the cluster in a set of point-charges
mimicking the rest of the solid. Such procedures are not well controlled,
e.g. depending on the details, the sign of crystal-field may change.\cite%
{Sugano63, Wachters71, Wachters72}

Here we use a different route to performing \textit{ab initio} MLFT
calculations. Our procedure is similar to the method originally devised by
Gunnarsson \textit{et al.} \cite{Gunnarsson89, Anisimov91} for obtaining
the parameters in the Anderson impurity model and, in the last 15 years,
used extensively for dynamical-mean-field calculations for realistic solids
(LDA+DMFT). \cite{Anisimov97, Georges96, Held03, Maier05, Lechermann06} We
start our \textit{ab initio} MLFT calculation by performing a DFT
calculation for the proper, infinite crystal using a modern DFT code which
employs an accurate density functional and basis set (e.g. LAPWs).\cite%
{Andersen75,Blaha90} From the (selfconsistent) DFT crystal potential we then
calculate \emph{a set of Wannier functions} suitable as single-particle
basis for the MLFT calculation.\cite{Andersen00, Andersen01, Andersen03, Pavarini05, Yamasaki06, Zurek05}
Since the members of such a set are centered either on the TM or ligand
atoms, we shall call them Wannier \emph{orbitals.} Typically, they are the
TM $3d$- and oxygen $2p$-orbitals which, \emph{taken together}, exactly
describe the DFT $\mathrm{3d}$- and $\mathrm{2p}$ bands. In general, the set should
be minimal and span \emph{exactly} all DFT solutions in the energy range
relevant for the property to be calculated. It is important that this set
contains sufficiently many ligand orbitals to make the \emph{correlated} TM 
\emph{orbitals well localized,} \textit{i.e.} the TM $d$-orbitals should not
have tails on any other atom. This localization allows one to restrict
the many-electron calculations of local properties to a single TM site plus
its ligand neighbors. Hence, in the current method, there are no embedding
errors, except those arising from truncating the single-particle basis to
include only the Wannier orbitals on the cluster.

In the following we introduce the method by the example of the late
transition-metal oxide NiO with configuration $3\mathrm{d}^{8}$. In Sect.
II, we show that similar results can be obtained for middle and early
transition-metal oxides, specifically $3\mathrm{d}^{5}$ MnO and $3\mathrm{d}%
^{0}$ SrTiO$_{3}.$ 

In Sect. III, we compare with results obtained by several different
experimental techniques: (A) $2p$ x-ray absorption (XAS), a charge neutral
excitation of a transition-metal $2p$ core electron into the $3d$ shell. (B) 
$2p$ core level x-ray photoemission (XPS) from Ni impurities in MgO. (C)
Inelastic x-ray scattering (IXS) of core to valence excitations, a technique
similar to XAS. We specifically show $3p$ core-electron excitations into the 
$3d$ shell. (D) Inelastic x-ray scattering of $d$-$d$ excitations. The
experiments presented for these materials are relatively well understood, so
that the comparison with our new \textit{ab initio} results constitute a
critical test of the theoretical method. At the end of the paper, we
conclude. In appendix \ref{appCovalence} we provide information on the different basis sets or Wannier orbitals used, as well as the meaning of the different occupation numbers and the concept of formal valence. Details of the calculations, including numerical values of several
MLFT parameters obtained \textit{ab initio,} may be found in Appendix \ref{appComp}. A
discussion of the double counting of interactions in the LDA and MLFT
calculations may be found in Appendix \ref{appDouble}. In appendix \ref{appBasisSize} we show how Ligand orbitals can be obtained in general symmetry from the O $2p$ orbitals, with the use of blocktridiagonalization of the orbital basis set. This is an essential ingredient which makes these calculation numerically efficient. Appendix \ref{appDiag} contains a short note on the exact diagonalization routines.

\section{Obtaining the MLFT parameters from the LDA by the example of N\lowercase{i}O}

In this section we introduce the \textit{ab initio} MLFT method by the example of NiO. We will discuss the different steps taken in order to obtain the MLFT parameters. First we discuss the LDA procedure used to obtain the potential, Wannier functions and tight binding parameters. Next we discuss the meaning of the different one electron parameters. In the last part of this section we discuss many body parameters, i.e. the Slater integrals.

We start our \textit{ab initio} calculations with a conventional
charge-selfconsistent LDA calculation for the experimental crystal
structure. NiO has the rock-salt structure in which each Ni atom is
surrounded by six O atoms in cubic symmetry, and vice versa. We used the
linear augmented plane wave (LAPW) method\cite{Andersen75,Singh00} as
implemented in \textsc{Wien2k.}\cite{Blaha90} The resulting LDA band
structure is shown along the symmetry lines of the face centered cubic (fcc) Brillouin zone in Fig.$\,$\ref{NiOLDA}. It is not very different from the band structure obtained
and discussed forty years ago by Mattheiss \cite{Mattheiss72} who used
Slater exchange and a non-selfconsistent potential construction. The three O 
$\mathrm{2p}$ bands extend over 5 eV, from -8.2 to -3.2 eV below the Fermi
level. The five Ni $\mathrm{3d}$ bands consist of three $\mathrm{t}_{2g}$
bands extending from -3.0 to -0.9 eV and two $\mathrm{e}_{g}$ bands
extending from -1.4 to +1.3 eV. The bottom of the Ni $\mathrm{4s}$ band is
1.5 eV above the Fermi level and at the $\Gamma $ point. As pointed out by
Mattheiss, the reason why the $\mathrm{4s}$ band is above the $\mathrm{3d}$
bands and thus empty, while it is half-full in elemental Ni, is that strong
hybridization with the O $\mathrm{2}p$ band pushes it up (and the $2p$ band
down) in the oxide.

Within the LDA, NiO is a metal, in strong contrast to experiments where NiO
is found to be a good insulator with a room-temperature resistance of $\sim
10^{5}\,\Omega $cm and an optical band-gap of about 3.0-3.5 eV. \cite%
{Morin54, Newman59, Powell70} This is one of the most noticeable failures of
the LDA. However, for the current paper, this is not a problem. Although the
LDA cannot reproduce the correct electronic structure near the nickel atom,
the minimal set of localized Ni $d$ and O $p$ orbitals which together span
the low-energy solutions of Schr\"{o}dinger's equation for the LDA crystal
potential\ exactly, i.e. the $5+3=8$ bands in Fig.$\,$\ref{NiOLDA}, is
expected to constitute a good single-particle basis set for many-body
calculations.

In order to prevent double counting of the multipole part of the Coulomb
interaction, we constrain the selfconsistent LDA potential to be spherically
symmetric inside non-overlapping muffin-tin (MT) spheres (see Appendices \ref{appComp}
and \ref{appDouble}), but allow it to be general in the MT-interstitial; it is a so-called 
\emph{warped} MT potential. For this potential we generate a basis set of 8
localized TM $d$ and O $p$ orbitals per cell which span the 8 bands exactly.
Since these bands do not cross any other bands in NiO, this can be done by
projection of the LDA LAPW Bloch states onto Wannier functions choosing
band- and $\mathbf{k}$-dependent phases which make the Wannier functions
atom-centered and localized. For an oxide like SrTiO$_{3}$, the TM $d$ and O 
$p$ bands \textit{do} cross and hybridize with other bands far away from the Fermi
level as can be seen in Fig.$\,$\ref{LDACompareNiOMnOSrTiO3}; near avoided
crossings it is therefore not clear which of the bands to project on to.
Moreover, one might want to go beyond perfect crystals. Rather then using
projection, we generate the minimal basis set of localized orbitals \emph{%
directly} by using the $N$th-order muffin-tin orbital ($N$MTO) method.\cite%
{Andersen00, Andersen01, Andersen03} This method solves the problem exactly
by multiple scattering theory at $N+1$ chosen energies, followed by $N$%
th-order polynomial interpolation in the Hilbert space, but only for a
superposition of spherically symmetric short-ranged potentials (to leading
order in the potential overlap). We must therefore first perform the
overlapping muffin-tin approximation (OMTA)\cite{Zwierzycki09} to the warped
MT potential, i.e. by least-squares minimization determine the radial shapes
of the overlapping potential wells and the common potential zero.

The resulting basis set of five Ni $d$ plus three O $p$ $N$MTOs\cite{NotLMTO, Andersen84}
per cell with the $N+1=2$ energies, $\epsilon _{0}=-5.2$ and $\epsilon
_{1}=-1.2$ eV, produces the eight colored, thick bands in Fig.$\,$\ref%
{NiOLDA}. Within the width of the line they are indistinguishable from the
LAPW bands. Hence, the $N$MTO minimal basis set for the OMTA to the warped
potential is a highly accurate representation of the large LAPW basis set
for the low-energy states, but many times more efficient. Our Wannier
orbitals are symmetrically orthonormalized $N$MTOs\cite{Andersen00,
Andersen01, Andersen03} and the colors indicate the relative O $p$, Ni $%
d\left( t_{2g}\right) ,$ and Ni $d\left( e_{g}\right) $ Wannier-orbital
characters. The middle panel of Fig.$\,$\ref{NiOLDA} shows the partial
density of states projected onto these Wannier orbitals. Compared with the
commonly used projection onto truncated partial-waves inside a MT sphere,
our projection has the advantage of leading to well-defined occupation
numbers because it is onto a complete, orthonormal basis set of localized,
smooth orbitals. Our projection also takes care of the O $p\left(
t_{1u}\right) $ character which flows into the neighboring Ni MT sphere due
to the large size of the Wannier O $p$ orbital. In this regard, it should be
remembered that a MT {sphere} is \emph{not} chosen to give a good
representation of the charge density, and hence of the occupied Wannier
orbitals, but of the \emph{potential}. Since the latter has an envelope
function which for rocksalt-structured NiO is essentially the Coulomb
potential from equal, but opposite point charges on identical cubic
sublattices, Ni and O have \emph{similar sized} MT spheres. This makes it
necessary for the wave-functions resulting from a MT-based method for
solving Schr\"{o}dinger's equation such as LAPW, to carry the partial-wave
expansions much further than to $p$ or $d$, typically to $l$ $\sim 8$,
because the outer part of the O $p$ orbitals, for instance, are being
expanded around the Ni sites. Nevertheless, with appropriately normalized
partial waves, projection of the density of states does give similar results
as with Wannier $pd$-orbitals. 

\begin{figure}[h]
\includegraphics[width=0.5\textwidth]{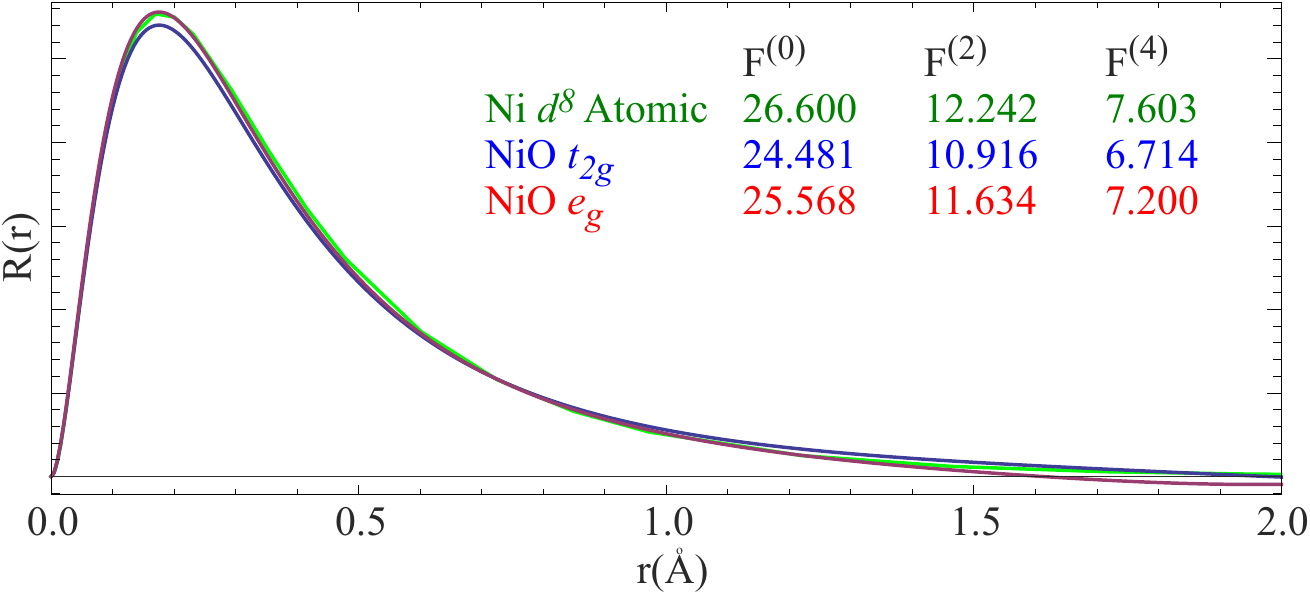}
\caption{(color online) Angular averaged radial wave functions, $R(r)$, for
the Ni $e_{g}$ and Ni $t_{2g}$ Wannier orbitals compared with the
Hartree-Fock radial wave function for a Ni$^{2+}$ ion in a $3d^{8}$
configuration. For Ni$^{+}$ $3d^{8}s^{1}$ and Ni $3d^{8}s^{2},$ the radial
functions are similar. The distance to the nearest oxygen is 2.09 {\AA },
which is consistent with the sum of the ionic radii of 0.72 {\AA } for Ni$%
^{2+} $ and 1.40 {\AA } for O$^{2-}.$ The inset shows the Slater integrals
Eq. (\protect\ref{Slater}) for the multipole Coulomb interactions.}
\label{NiORofr}
\end{figure}

The eight Wannier orbitals, $w_{i}\left( \mathbf{r}\right) $, are shown on
the right-hand side of Fig.$\,$\ref{NiOLDA} as those surfaces where $%
\left\vert w_{i}\left( \mathbf{r}\right) \right\vert =\mathrm{const}$ and
which incorporate 90\% of the charge, $\int_{S}\left\vert w_{i}\left( 
\mathbf{r}\right) \right\vert ^{2}d^{3}r\equiv 0.9.$ The red/blue color of a
lobe gives its sign. As one can see, the Ni $d$ orbitals are extremely well
localized. This is a necessary condition for several many-body models which
implicitly assume such an orbital basis set, for example the Hubbard model
which neglects all off-site Coulomb correlations. In order to visualize the
localization of the Ni 3$d$ Wannier orbitals at a more quantitative level,
we computed the effective radial wave-functions for the $t_{2g}$ and $e_{g}$
orbitals by multiplying with the corresponding spherical harmonics and
averaging over all solid angles. These radial functions are compared in Fig.$%
\,$\ref{NiORofr} with that of a Ni atom in the $d^{8}$ configuration as
calculated with the Hartree Fock method.\cite{Cowan81} Although there are
slight differences, the agreement is astonishing. The local Ni $d$ Wannier
orbitals in NiO are rather similar to atomic Ni wave-functions. Note that
the atomic Ni $d$ radial function depends on the filling of the $d$-shell,
but is rather insensitive to the filling of the $4s$ shell. The atomic
radial function shown in Fig.$\,$\ref{NiORofr} is calculated for a Ni$^{2+}$ 
$\left( 3d^{8}4s^{0}\right) $ configuration, but would be practically the
same for a Ni$^{+}$ $\left( 3d^{8}4s^{1}\right) $ or neutral Ni $\left(
3d^{8}4s^{2}\right) $ configuration.\bigskip

Since we have chosen not to include Ni $s$ orbitals in the minimal basis, it
does not describe the high-lying, empty Ni $\mathrm{4s}$ band which has
anti-bonding O 2$p$ character. The corresponding Ni $4s$ bonding character
of the O 2$p$-like band is however completely taken care of by including
(downfolding) the Ni 4$s$ character into the tails O 2$p$ Wannier orbitals,
as is seen in Fig.$\,$\ref{NiOLDA}. In the bottom right-hand panel one can
see how Ni $4s$ character is added at the tip of each lobe of the O $2p$
orbital, such that the outermost $4s$ radial lobe expands the tip of the $2p$
lobe, while the remaining inner radial $4s$ lobes of alternating sign cause
the $2p$ lobe to tail-off in an oscillating manner. The shape of the O $2p$
Wannier orbital is of course also influenced by the requirement that it be
orthogonal to the Ni 3$d$ Wannier orbitals.

The $N$MTO method is particularly useful when a real-space tight-binding
representation of the Hamiltonian is needed.\cite%
{Gunnarsson89, Anisimov91,Anisimov97,Lechermann06,Zurek05,Pavarini05} Both
the orthogonal Wannier functions as well as the corresponding tight-binding
representation of the Hamiltonian in this basis set are directly available
in the $N$MTO formalism. Details on the $N$MTO method can be found in previous publications\cite{Andersen00, Andersen01, Andersen03} and Appendix \ref{appComp}. Here we would like to stress that the Wannier orbitals used within this paper are not constructed by maximally localizing the Wannier functions,\cite{Kunes10} but their extend is a result of symmetry constraints. This leads to orbitals that are not always maximally localized, especially in the details of the tails of these orbitals. The Ni $d$ Wannier orbitals obtained by $N$MTO do show a very large overlap with atomic orbitals, which might well be larger than the overlap one might find between atomic orbitals and maximally localized Wannier orbitals. It is the agreement between our Wannier orbitals and atomic orbitals which makes the method successful. An alternative method to obtain good Wannier orbitals for correlated model calculations could be to maximize the overlap of the Wannier orbital with atomic orbitals. 

Although only the Ni $\mathrm{d}$ bands in Fig.$\,$\ref{NiOLDA} are partly
occupied, inclusion of O $p$ orbitals in the basis is important for
describing spectroscopy. In photoemission, for example, the removal of a TM $%
d$ electron can lead to a transfer of charge from the O to the TM atom. This
dynamical screening would not be captured on a basis of only TM $\mathrm{d}$ orbitals. Multiplet \emph{Crystal} Field Theory
(MCFT), i.e. local calculations using a basis of only TM $\mathrm{d}$
orbitals, which have antibonding O $p$ tails fixed to them, can be
useful in many other cases, for example for calculating magnetic
anisotropies. In this paper, however, we concentrate on Multiplet \emph{%
Ligand }Field Theory (MLFT) and explicitly include the O $p$ orbitals in the
basis set.

\begin{figure*}[t]
\includegraphics[width=0.95\textwidth]{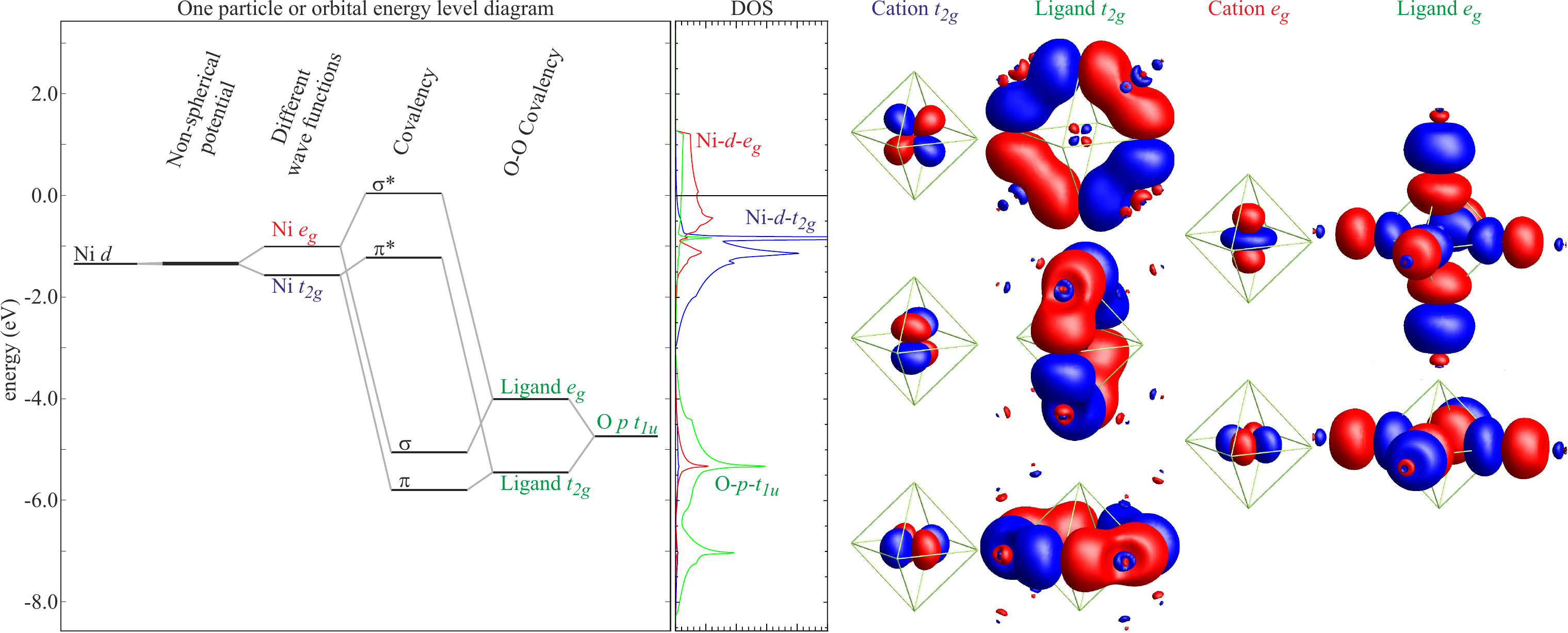}
\caption{(color online) Left panel: Orbital energy level diagram for the NiO$%
_{6}$ cluster on the same energy scale as the LDA band-structure for the
solid (Fig.$\,$\protect\ref{NiOLDA}) shown in the middle panel. The Fermi
level is the zero of energy. Right panel: Constant-amplitude contours of the
Ni $d$ Wannier orbitals and of the Ni-centered ligand orbitals. The latter
are symmetrized linear combinations of the O $p$ Wannier orbitals.}
\label{NiOClusterWanierOrbitalEnergy}
\end{figure*}

Until now, we have considered the infinite crystal and have calculated the
one-electron potential in the local-density and warped muffin-tin
approximations. For that potential we have derived a set of localized O $p$
and Ni $d$ Wannier orbitals which exactly describe the O $\mathrm{p}$ and Ni $\mathrm{d}$ bands, as well as the corresponding tight-binding Hamiltonian.
We now use these orbitals and this Hamiltonian for the NiO$_{6}$ cluster (Fig.$\,$\ref{NiO6Cluster}). The
band structure thereby reduces to the O $p$ like $\pi $- and $\sigma $%
-levels and the Ni $d$ like $\pi ^{\ast }$- and $\sigma ^{\ast }$-levels
shown in the central part of the left-hand panel of Fig.$\,$\ref%
{NiOClusterWanierOrbitalEnergy}, plus some O $p$ levels which do not couple
to the Ni $d$ levels and are therefore not shown in the figure. For
comparison, we repeat from Fig.$\,$\ref{NiOLDA} the crystalline density of
states projected onto the O $p,$ Ni $d\left( t_{2g}\right) $ and $d\left(
e_{g}\right) $ orbitals. In the following we discuss the formation of these
simple cluster levels before we consider calculating many-electron
multiplets.

The NMTO Ni $d$ Wannier orbitals have the on-site energies $\epsilon
_{t_{2g}}=-$1.55 and $\epsilon _{e_{g}}=-1.05\,$eV with respect to the Fermi
level. Had the potential been spherically symmetric within the range of the
Ni $d$ orbitals, the $e_{g}$ and $t_{2g}$ radial functions in Fig.$\,$\ref%
{NiORofr} would have been identical, and their levels degenerate with energy 
$\epsilon _{d}=\frac{3}{5}\epsilon _{t_{2g}}+\frac{2}{5}\epsilon _{e_{g}}=-$%
1.35 eV. The crystal-field splitting, $10Dq=\epsilon _{e_{g}}-\epsilon
_{t_{2g}}=0.5\,$eV, is basically due to the fact that $e_{g}$ and $t_{2g}$
orbitals point respectively towards and between the nearest oxygen
neighbors, which are negatively charged. (The notation $10Dq$ for $\epsilon_{e_{g}}-\epsilon _{t_{2g}}$ is standard in MLFT.\cite{10Dq}) In the conventional ionic picture,
2 electrons are transferred from each neutral Ni $3d^{8}4s^{2}$ atom to each
O atom, where they complete the $2p$ shell. Hence, the crystal-field
splitting would be the radial matrix element of the non-spherical part of the
electrostatic Madelung potential $\propto r^{4}\left[ Y_{40}\left( \mathbf{%
\hat{r}}\right) +\sqrt{(}5/14)[Y_{44}\left( \mathbf{\hat{r}}\right)
+Y_{4-4}\left( \mathbf{\hat{r}}\right) ]\right] $ from these $\pm 2$ 
charges. However, with the atomic radial function shown in Fig. \ref{NiORofr}, which yields: $\left\langle r^{4}\right\rangle \approx \left(0.71\,\mathrm{\mathring{A}}\right) ^{4}$, the splitting is merely $\sim
0.3\,$eV, \textit{i.e.} $\sim 0.2\,$eV too small, and this is even an
overestimate because the charge transfer from the $\mathrm{4s}$ to the $%
\mathrm{2p}$ band is reduced by covalency. Note in passing, that had we
taken the anisotropy of the electrostatic potential to be the one caused by
the LDA charge density and the protons outside the Ni MT sphere, we would
have gotten merely 0.01\thinspace eV. This is so because the Ni MT radius of
1.10 {\AA} is larger than the Ni$^{2+}$ ionic radius of 0.72 {\AA} and
thus cuts off part of the oxygen charge density (remember: MT spheres are
designed to describe the potential and not the charge density). Hence, the
anisotropy felt by the different \emph{angular} behaviors of the $e_{g}$ and 
$t_{2g}$ orbitals can at most only account for half the calculated
crystal-field splitting. Next, we now turn to the different \emph{radial}
behaviors (Fig.$\,$\ref{NiORofr}). These are mostly due to the requirement
that the Ni $e_{g}$ and $t_{2g}$ Wannier orbitals be orthogonal to the
nearest O $2p$ orbitals. The $e_{g}$ radial function changes sign for
increasing $r$, while the $t_{2g}$ radial function merely decays. At short
distances ($r\lesssim 0.7$ {\AA }), the normalized $e_{g}$ radial function
is therefore larger than the normalized $t_{2g}$ function. Since the maxima
of the two radial functions occur where the radial potential, $v_{\mathrm{Ni}%
}\left( r\right) +6/r^{2},$ is huge and negative, the higher $e_{g}$ maximum
causes a lower potential energy, opposite to what is needed to explain the
size of the crystal-field splitting. In the end, it turns out that the $\sim$0.5 eV crystal-field splitting is a result of not only the potential acting on the different angular and radial wave-functions, but also due to the \emph{kinetic} energy. The $e_{g}$ orbitals overlap more with the O $p$ orbitals than the $t_{2g}$ orbitals, whereby orthogonalization increases the kinetic energy more for the
former than for the latter. For the calculation of the MLFT parameters it turns out to be important to treat all these interactions on an equal footing. 

We now continue our explanation of the orbital energy level diagram in Fig. \ref{NiOClusterWanierOrbitalEnergy}, this time starting from on-site
energy $\epsilon_{p}=-4.74$ eV of the O $2p$ Wannier orbitals marked on the
right-hand side. The ionic energy is thus $\epsilon_{d}-\epsilon
_{p}\approx 3.4$ eV. Since we have chosen not to include Ni $4s$ orbitals in
the basis set, the bonding Ni $s$ character has been downfolded into the O $p
$ orbitals so that the $-4.75$ eV includes a downwards shift of about $1$ eV
from Ni $s$ covalency. The NiO$_{6}$ cluster contains $6\times 3=18$ O $p$
orbitals, but not all linear combinations can interact with $d$ orbitals on
the central Ni site. Hence, the basis set for the MLFT calculations can be
greatly reduced by including only those linear combinations which do couple,
the so-called ligand ($L$) orbitals.\cite{Ballhausen62} The reduction of the Hilbert space by use of Ligand orbitals is crucial for efficient MLFT calculations which is explained in more detail in Appendix \ref{appBasisSize}. For each TM $d$
orbital there is exactly one such linear combination. The right-hand panel
of Fig. \ref{NiOClusterWanierOrbitalEnergy} shows the 5 Ni $d$ orbitals
together with the 5 Ni-centered $L$ orbitals of the same symmetry. There is
an important difference between the $L$ $t_{2g}$ and $L$ $e_{g}$ orbitals:
Whereas the $L$ $t_{2g}$ orbitals are bonding (same color) between nearest O 
$p$ Wannier orbitals, and thus give rise to a substantial O-O $\sigma $-like
bond charge, the $L$ $e_{g}$ orbitals are anti-bonding (different color). As
a result, the energies of the $L$ $t_{2g}$ and $e_{g}$ orbitals are
respectively $T_{pp}=pp\sigma -pp\pi $ below and above $\epsilon _{p}$.\cite%
{Ghijsen88,Eskes90}

We finally complete the level diagram by including the covalent hopping
integrals $V_{t_{2g}}=pd\pi \times 4/\sqrt{4}=2pd\pi $ and $V_{e_{g}}=-\frac{%
\sqrt{3}}{2}pd\sigma \times 4/\sqrt{4}=-\sqrt{3}pd\sigma \sim 3pd\pi $\cite%
{Andersen78} between the $L$ $p$ and TM $d$ orbitals of respectively $t_{2g}$
and $e_{g}$ symmetry. The $t_{2g}$ hopping gives rise to an $L$ $p$-like $%
\pi $ and a TM $d$-like $\pi ^{\ast }$ level, and the $e_{g}$ hopping gives
rise to an $L$ $p$-like $\sigma $ and a TM $d$-like $\sigma ^{\ast }$ level.
It is these $\pi ^{\ast }$ and $\sigma ^{\ast }$ levels which in the solid
broaden into Ni $\mathrm{t}_{2g}$ and $\mathrm{e}_{g}$ bands. The $\sigma
^{\ast }$ level is close to the Fermi level in the LDA and this indicates
that the $\sigma ^{\ast }$ orbital is half full. The $\pi $, $\sigma $, and $%
\pi ^{\ast }$ orbitals have considerably lower energies and are fully
occupied.

Our MLFT calculations include Coulomb correlations beyond the one-electron
mean-field potentials discussed so-far, but only among the TM $d$ orbitals.
Arguments for treating the $L$ $p$ orbitals as well as their Coulomb
repulsion with the TM $d$ orbitals at the LDA level, are that the $L$ $p$
orbitals are fairly delocalized and that they are almost fully occupied. As
an example, we can safely neglect correlation in an event where two holes
meet on a single oxygen atom and scatter. The Coulomb correlations are
responsible for the multiplet structure, and we keep them among the Ni $d$
orbitals, but make a distinction between the spherical ($U$, $\Delta $) and
the non-spherical repulsions.

The spherical part of the Coulomb repulsion, often parametrized by $U$, is
strongly screened in a solid. If a Ni $d$ electron is removed, there will be
a charge-flow into the Ni $4s$ orbital, for example, which reduces the
energy cost of such an excitation. Although several calculations of the
screened $U$ have been presented in the past,\cite{McMahan88, Hybertsen89, Anisimov91, Springer98, Aryasetiawan04, Anisimov05, Cococcioni05, Cortes07, Anisimov09, Miyake10, Sasioglu11, Franchini11} we fit $U$ such as to obtain the
best agreement between our MLFT calculation and the experimental multiplet
spectra.\cite{Bocquet96,Tanaka94} The parameter $\Delta $ is the orbitally
averaged (spherical) part of the difference between the on-site energies of
the Ni $d$ orbitals and the $L$ $p$ orbitals at a filling of $8$ electrons in the Ni $d$ shell. In the LDA, as well as for the
ground-state found in our MLFT calculations, the Ni $d$ occupation exceeds $8$ due to the covalency with the oxygens. The relation between $\Delta$,
as defined in MLFT calculations, and $\epsilon _{d}-\epsilon _{p},$ as
obtained from the LDA, is rather non-trivial and we shall therefore treat,
not only $U$, but also $\Delta$ as an adjustable parameter. In the
foreseeable future, it should be possible to calculate $U$ and $\Delta $
from first principles.

The non-spherical parts of the Coulomb interactions we can easily calculate
because the multipole interactions between two $d$ electrons are hardly
screened. For example, the Coulomb repulsion between two $d_{x^{2}-y^{2}}$
electrons is obviously larger than that between a $d_{x^{2}-y^{2}}$ electron
and a $d_{3z^{2}-1}$ electron, but to screen this difference requires
electrons with high angular momentum around the Ni site; a Ni $4s$ electron,
for instance, could \emph{not} do it. Also electrons on neighboring sites
are inefficient in screening the multipole because it decays fast ($\propto
r^{-k-1}).$ {It has been shown that neglecting any screening of the
multipole part of the Coulomb interaction gives reasonable agreement between
theory and experiment.\cite{Antonides77}} Also in the present work, we shall
neglect any screening of the multipole part of the Coulomb interaction and
shall find reasonable agreement with experiments.

Multipole interactions are the cause of the Hunds-rule energy. For example,
two $d_{x^{2}-y^{2}}$ electrons must have different spins, whereas two
electrons in different $d$ orbitals, and hence less repulsive, may be in a
spin-triplet state, as well as in the spin-singlet state. Experimentally it
has been shown \cite{Antonides77} that the multipole interactions of the
Coulomb interaction, are reasonably well approximated by assuming that the $%
d $ orbitals have the pure-angular-momentum form: $R\left( r\right)
Y_{2m}\left( \mathbf{\hat{r}}\right) $. The inset in Fig.$\,$\ref{NiORofr}
is a table of the values of the Slater integrals obtained using the Ni$^{2+}$
ionic radial function, $R\left( r\right) $, as well as the radial functions
obtained by averaging the Ni $t_{2g}$ and $e_{g}$ Wannier orbitals over
solid angles. The Slater integrals for $d$ orbitals are: 
\begin{equation}
F^{(k)}=\int \int \frac{r_{<}^{k}}{r_{>}^{k+1}}R^{2}(r_{1})R^{2}(r_{2})%
\,r_{1}^{2}dr_{1}\,r_{2}^{2}dr_{2}.  \label{Slater}
\end{equation}%
where $r_{<}=\min (r_{1},r_{2}),$ $r_{>}=\max (r_{1},r_{2}),$ and $k=0,2,$or
4. The definitions of $U$ and the Hund's rule exchange, $J_{H},$ vary: The
average repulsion between two $d$ orbitals is: $U_{av}=F^{(0)}-\frac{14}{441}%
(F^{(2)}+F^{(4)})$. However, in order to discuss the Mott gap, one uses the
energy difference between the lowest multiplets of different configurations
and that has lead to the definition: $U=F^{(0)}+\frac{4}{49}F^{(2)}+\frac{36%
}{441}F^{(4)}$. The Hund's-rule exchange can either be defined as: $%
J_{H}=\frac{1}{14}(F^{(2)}+F^{(4)}),$ or as: $J_{H}=\frac{2.5}{49}F^{(2)}+%
\frac{22.5}{441}F^{(4)}$. The bare $F^{(0)}$ as calculated from the Wannier
orbitals is of the order of $\sim $25 eV. This is clearly much too large
because the monopole part of the Coulomb repulsion is strongly screened. The
values of $F^{(2)}$ and $F^{(4)}$ are respectively $\sim 11$ and $\sim 7$
eV, in good agreement with experimental values, as we shall see. The
multiplet interactions are quite large and lead to a multiplet splitting of
the Ni-$\mathrm{d}^{8}$ configuration of about 7.5 eV, which is the energy
difference between the $^{3}F$ ground-state configuration and the highest
excited singlet of $^{1}S$ character. This is larger than the Ni-$\mathrm{d}$
bandwidth and therefore \emph{not} a small energy.

We will compare our results to several experiments, including core level
spectroscopy. Once a core hole is made, the interaction between the core and
valence electrons becomes important. Here again we will make a distinction
between the multi- and monopole interactions. The monopole interactions $%
U_{2p,3d}$ and $U_{3p,3d}$ will, like for the valence states, be taken
from fits to experiment.\cite{Bocquet96,Tanaka94} For the multipole
interactions, \textit{i.e.} the Slater integrals $F_{p,d}^{2}$, $G_{p,d}^{1}$
and $G_{p,d}^{3},$ we again assume that screening can be neglected, which
allows us to directly calculate these integrals from the core and valence
Wannier orbitals. The core Wannier orbitals are equivalent to atomic wave
functions since they have no inter-site overlap. It is important to use a
scalar-relativistic method for the calculation, as well as to calculate the
core wave-functions for the final state occupations, which requires an open
shell calculation. We used an atomic Hartree-Fock code to obtain these core
wave-functions,\cite{Cowan81} but any open-shell, scalar relativistic method
should give similar results. Specific values of the Slater integrals can be
found in Table \ref{ParameterTable} in Appendix \ref{appDouble}. 

We now have all ingredients needed to perform MLFT calculations of
experimentally observable quantities. But before we do this, we will
introduce similar ligand-field calculations for MnO and SrTiO$_{3}$. This
allows us to compare oxides of early, intermediate and late transition
metals and show that the method is likely to apply to a range of correlated
transition-metal compounds.

\section{N\lowercase{i}O, M\lowercase{n}O and S\lowercase{r}T\lowercase{i}O$%
_{3}$}

\begin{figure}[tbp]
\includegraphics[width=0.5\textwidth]{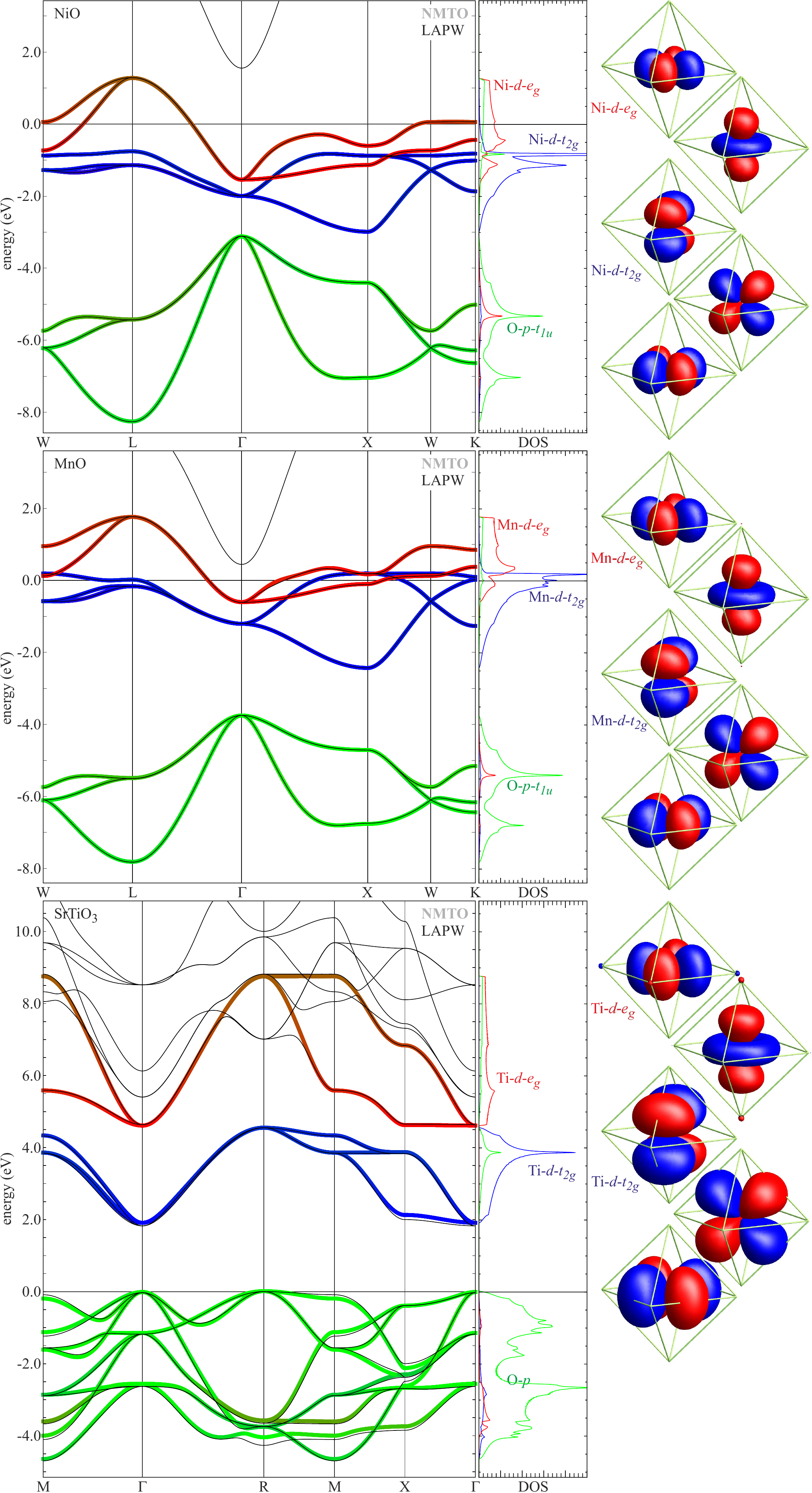}
\caption{(color online) Same as Fig.$\,$\protect\ref{NiOLDA}, but for fcc
NiO $\left( \mathrm{d}^{8}\right) $, fcc MnO $\left( \mathrm{d}^{5}\right) $%
, and sc SrTiO$_{3}$ $\left( \mathrm{d}^{0}\right) $. On the right-hand
side, the oxygen orbitals are not shown.}
\label{LDACompareNiOMnOSrTiO3}
\end{figure}

\begin{figure}[h]
\includegraphics[width=0.5\textwidth]{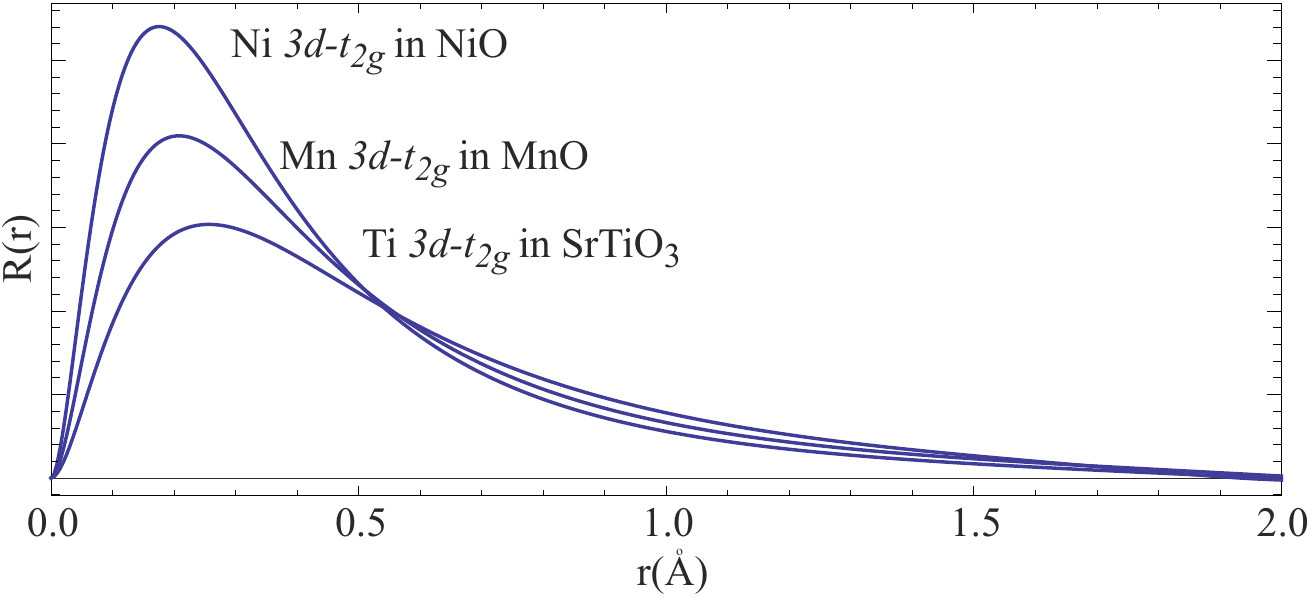}
\caption{(color online) Angular averaged TM $t_{2g}$ Wannier-orbitals in
NiO, MnO, and SrTiO$_{3}$. The distance to oxygen is 2.09 in NiO, 2.21 {\AA }
in MnO, but only 1.95 {\AA } in SrTiO$_{3}.$The ionic radii are 0.72 for Ni$%
^{2+}$, 0.80 for Mn$^{2+}$, 0.90\ for Ti$^{2+},$ but only 0.68 {\AA } for Ti$%
^{4+}$. }
\label{NiOMnOSrTiO3Rofr}
\end{figure}

In Fig.$\,$\ref{LDACompareNiOMnOSrTiO3} we shown from top to bottom the LDA
band-structures of NiO, MnO, and SrTiO$_{3}$ calculated in the same way as
in Fig.$\,$\ref{NiOLDA} and explained in the previous section, with details
given in Appendix \ref{appDouble}. Whereas NiO and MnO have the fcc rocksalt structure,
SrTiO$_{3}$ has the simple cubic (sc) perovskite structure in which the Sr cube has Ti at
its body center and O at its face centers; in MLFT we treat the TiO$_{6}$
cluster. Going from NiO to MnO, the TM-electron and -proton counts are both
reduced by 3, whereby the $\mathrm{d}$-band filling gets reduced from $%
\mathrm{d}^{8}$ to $\mathrm{d}^{5}.$ Concomitantly, we see that the $\mathrm{%
3d}$ bands move up in energy relatively to the $\mathrm{4s}$ and O $\mathrm{%
2p}$ bands, by about $1.5$ eV. The $\mathrm{p}$ and $\mathrm{e}_{g}$
bandwidths as well as the $\mathrm{e}_{g}$\textrm{-}$\mathrm{t}_{2g}$
splitting decrease, presumably due to the increased ionicity, $\epsilon
_{d}-\epsilon _{p}.$ Going finally to SrTiO$_{3},$ the TM-electron and
-proton counts are further reduced by 3, but due to the change of
stoichiometry, the nominal $\mathrm{d}$-band filling is now reduced to $%
\mathrm{d}^{0}$ rather than to $\mathrm{d}^{2}.$ SrTiO$_{3}$ is a band
insulator and the LDA bandstructure shown in the bottom panel agrees with
the ionic configuration Sr$^{2+}$Ti$^{4+}$(O$^{2-}$)$_{3}:$ We see nine full
O $\mathrm{2p}$ bands separated by a 2 eV gap from the three empty Ti $%
\mathrm{3d}\left( \mathrm{t}_{2g}\right) $ bands. The latter are separated
by a small gap from the two Ti $\mathrm{3d}\left( \mathrm{e}_{g}\right) $
bands which overlap the two Sr $\mathrm{4d}\left( \mathrm{e}_{g}\right) $
bands and the bottom of the Ti $\mathrm{4s}$ band. The three Sr $\mathrm{4d}%
\left( \mathrm{t}_{2g}\right) $ bands are pushed up in energy by covalent $%
pd\sigma $ interaction with the 12 nearest oxygen neighbors\cite{Pavarini05}
and thus lie more than 8 eV above the Fermi level. Due to the different
structure and stoichiometry of SrTiO$_{3},$ its bands are quite different
from those of NiO and MnO.

The agreement between the O $p$-like and TM $d$-like bands obtained with the
LAPW method and those obtained with the minimal basis set of $N$MTOs is
almost perfect for NiO and MnO. The agreement is also satisfactory for SrTiO$%
_{3},$ although near the bottom of the Ti $\mathrm{t}_{2g}$ band, and at
various places in the O \textrm{2}$\mathrm{p}$ band, small discrepancies may
be detected. These are most likely due to the OMTA causing a slightly
inaccurate description of the hybridization with the high-lying Sr $4d$ and
Ti $4s$ bands.

From the right-hand side of the figure, we see that for all three materials
the TM $3d$ Wannier orbitals are very well localized. This is a necessary
condition for using them in MLFT. We do not show the O $2p$ orbitals as in
Fig.$\,$\ref{NiOLDA}, but had we done so for SrTiO$_{3},$ we would have seen
not only bonding Ti $4s$ character of the $p$ orbital pointing towards Ti,
but also bits of bonding Sr $4d$ and $5s$ characters of the two other $p$
orbitals, which point towards Sr.\cite{Pavarini05} The good localization of the Ti $e_{g}$
orbitals is related to the feature seen in the left-hand panel around 8 eV,
that the $N$MTO Ti\textrm{\ }$\mathrm{e}_{g}$ band interpolates \emph{%
smoothly} across the avoided crossing of the LAPW Ti $\mathrm{e}_{g}$ and Sr 
$\mathrm{e}_{g}$ bands. Had this not been the case, the Ti $e_{g}$ Wannier
orbitals would have been long-ranged. Hence, we can construct a minimal set
of \emph{localized} $N$MTOs, even when bands described by the set are
crossed by and hybridizes with other bands, provided that we can choose the $%
N+1$ expansion energies outside the range of those other bands. For SrTiO$%
_{3},$ we used $\epsilon _{0}=-2.6$ and $\epsilon _{1}=1.5$ eV.

The $3d$ Wannier orbitals for the three oxides are very similar; they merely
contract along the $3d$ row of the periodic table. This is seen when going
from the bottom to the top in the right-hand panel of Fig.$\,$\ref%
{LDACompareNiOMnOSrTiO3}, and even more clearly in Fig.$\,$\ref%
{NiOMnOSrTiO3Rofr} where we show the angular-averaged $t_{2g}$ Wannier
orbitals. The well-known reason for this orbital contraction is as follows:
Upon proceeding one step along the TM row, a proton and an electron are
added. The electron incompletely screens the attractive potential from the
proton seen by another valence electron, and as a result, the one-electron
potential becomes deeper and deeper upon proceeding along the series, until
the $d$ shell is full and the screening is complete. Since this mechanism is
atomic, the shapes of our $3d$ Wannier functions are fairly robust and the
chemistry merely changes tails --and thereby normalizations-- a bit. This is
what we saw in Fig.$\,$\ref{NiORofr}. For that reason, the contraction seen
in Fig.$\,$\ref{NiOMnOSrTiO3Rofr} of the $t_{2g}$ radial functions --which
are less influenced by O than the $e_{g}$ functions-- closely follows that
of the 2+ ionic radii, which are 0.72 (Ni$^{2+}$), 0.80 (Mn$^{2+}$), and
0.90 {\AA } (Ti$^{2+}$), in the sense that at the respective radius, all
three radial functions have about the same amplitude. This happens although
the Wannier orbitals are calculated for the real solids, which in the case
of SrTiO$_{3}$ have a Ti-O distance far smaller than expected from the size
of the Ti $t_{2g}$ function in Fig.$\,$\ref{NiOMnOSrTiO3Rofr}. That the
chemical binding of SrTiO$_{3}$ is different than those of NiO and MnO also
becomes clear by adding the accepted ionic radius of O$^{2-}$ (1.40 {\AA )}
to the TM$^{2+}$ radii given above, obtaining: 2.12, 2.20, and 2.30 {\AA }
for the TM-O distance in respectively NiO, MnO, and SrTiO$_{3}.$ The real
distances in NiO and MnO are nearly the same, but much smaller (1.95 {\AA })
in SrTiO$_{3}.$

The short Ti-O distance is of course reproduced by using the Ti$^{4+}$
radius of 0.68 {\AA }\ which corresponds to the band-structure configuration
Sr$^{2+}\,$Ti$^{4+}\,$(O$^{2-}$)$_{3}=$ Sr$\,\mathrm{4d}^{0}$\ Ti$\,3\mathrm{%
d}^{0}\,$(O$\,\mathrm{2p}^{6}$)$_{3}.$ This ionic picture of the binding
seems to neglect the Ti-O and Sr-O covalencies predicted by the LDA, \textit{%
i.e.} the fact that there is a considerable amount of Ti-$3d$ and Sr-$4d$
partial-wave character in the O $\mathrm{2p}$ bands. But this is only
apparently so: The Ti $3d$ and Sr $4d$ radial functions are essentially
exponentially increasing because they solve the respective radial Schr\"{o}%
dinger equation for O $\mathrm{2p}$-band energies, which are way below those
of the Ti $\mathrm{3d}$ and Sr $\mathrm{4d}$ bands. Hence, these partial waves
simply complete the shape of the O $\mathrm{2p}$ Wannier orbitals inside the
Ti and Sr MT spheres.

\section{Comparison with experiments}

In order to test the quality of MLFT calculations using the LDA Wannier
orbitals, we now present a comparison between theory and experiment for
several established spectroscopies, which show TM excitons. Such locally
bound states are represented well within the small cluster used in the MLFT.
The materials considered, namely, NiO, MnO and SrTiO$_{3}$, are insulators,
thus justifying the theoretical methodology further. NiO, MnO and SrTiO$_{3}$
have local ground states which are well understood, and the spectra shown
here have already been explained in the literature. New in the present paper
is that the MLFT parameters (except for $U$ and $\Delta $) are not fitted to
the experiment, but calculated \textit{ab initio}.

In the following subsections we first discuss x-ray absorption (XAS) at the $%
L_{2,3}$ edge, which probes TM $2p$ to $3d$ excitations. Next we show TM $2p$
core-level x-ray photoemission (XPS) experiments on an impurity system. Both 
$2p$ XAS and $2p$ XPS excite the same core states and the difference is that
in x-ray absorption the electron is excited into the local $3d$-shell
whereas photoemission excites the core electron into vacuum states. The
resulting spectra are very different. We then continue with core excitations
measured with inelastic x-ray scattering (IXS). From a theoretical point of
view, inelastic x-ray scattering of core to valence excitations and x-ray
absorption of core to valence excitations is very similar. The initial and
final states probed are the same. The difference is that whereas x-ray
absorption is mainly caused by dipole transitions, inelastic x-ray
diffraction is caused by multipole transitions determined by the length of
the transferred momentum. Finally in the last subsection, we show inelastic
x-ray scattering of $d$-$d$ excitations in NiO. These spectra are
particularly instructive as they allow for a relative straight-forward
understanding on how the different interactions contribute to each multiplet
excitation. For pedagogical reasons we provide a brief introduction to each
of the experimental techniques. A more thoroughly description of these
techniques can be found in textbooks, e.g. those by De Groot and Kotani, 
\cite{Groot08} St{\"{o}}hr\cite{Stohr92}, and Sch{\"{u}}lke.\cite{Schulke07}

\subsection{\textit{L}$_{2,3}$ edge x-ray absorption}

\begin{figure}[tbp]
\includegraphics[width=0.5\textwidth]{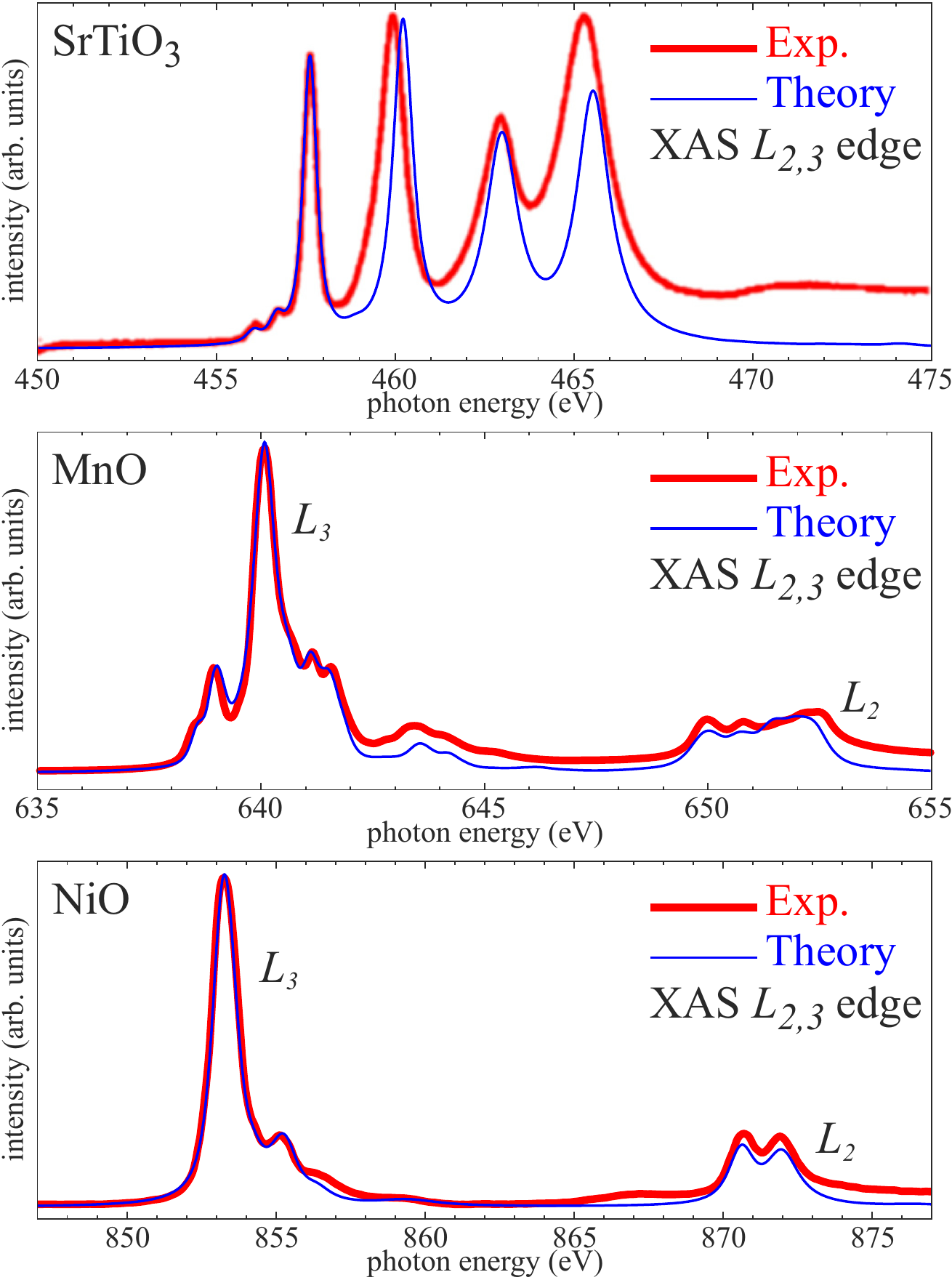}
\caption{(color online) Comparison of the experimental (thick red) and MLFT
(thin blue) TM $2p$ ($L_{2,3}$ edge) core-level x-ray absorption spectra for
SrTiO$_{3}$, MnO, and NiO. The experimental SrTiO$_{3}$ spectra are
reproduced from Uehara \textit{et al.}., \protect\cite{Uehara97} the MnO
spectra from Csiszar \textit{et al.,} \protect\cite{Csiszar05A,Csiszar05B}
and the NiO spectra from D. Alders \textit{et al.}.\protect\cite{Alders98}}
\label{ExpL23XAS}
\end{figure}

X-ray absorption spectroscopy (XAS) at the TM $L_{2,3}$ edge is
a technique whereby a TM $2p$ core electron is excited into the $3d$ valence
shell. The excitation energy is in the x-ray range and varies from $\sim 440$
eV for Ti to $\sim 855$ eV for Ni. The excitations are dipole allowed, which
make them so intense that spectra with very little noise can be obtained.
The spectra split into two set of peaks, the $L_{3}$ and $L_{2}$ edges, due
to spin-orbit coupling in the TM $2p$ core hole. This results in a $%
2p_{j=3/2}$ level ($L_{3}$ edge) lying $\frac{3}{2}\zeta _{2p}$ above a $%
2p_{j=1/2}$ level ($L_{2}$ edge). For core levels the relativistic
spin-orbit coupling is strong and element dependent: $\zeta _{2p}$=3.78 eV for Ti, 6.85 eV for Mn, and 11.50 eV for Ni. Hence, in Fig. \ref{ExpL23XAS} we see the $L_{3}$ and $L_{2}$ edges at 640 and 650 eV for MnO, and at 852 and 870 eV for NiO.

For SrTiO$_{3},$ $\frac{3}{2}\zeta _{2p}$ is of similar size as the $2p$-$%
3d$ multiplet splitting, \textit{i.e.} the $L_{2}$ and $L_{3}$ edges
overlap. The splitting within an $L_{2}$ or $L_{3}$ edge is due to the
combined interaction of covalent ligand-field effects and Coulomb
interactions between the $3d$ electrons and between the $2p$ core hole and
the $3d$ electrons. This leads to the relatively involved spectra with many
features as seen in Fig.$\,$\ref{ExpL23XAS}. Even for SrTiO$_{3}$ where one
might be tempted to relate the four intense peaks in the $2p$ XAS spectrum
to excitations from the $2p_{j=3/2}$ or $2p_{j=1/2}$ orbitals into either
the $t_{2g}$ or the $e_{g}$ orbitals, the intensity ratios (${4:2}${\
between excitations from respectively $2p_{j=3/2}$ and $2p_{j=1/2}$ core
holes, and }${3:2}${\ between excitations to respectively $t_{2g}$ and $e_{g}
$ states) do not follow this one electron picture: Assigning the peaks at
458, 460, 463 and 465 eV to excitations of the form $2p_{j=3/2}\rightarrow
t_{2g}$, $2p_{j=3/2}\rightarrow e_{g}$, $2p_{j=1/2}\rightarrow t_{2g}$, and $%
2p_{j=1/2}\rightarrow e_{g}$ respectively, would yield the intensity ratios }%
${12:8:6:4}$, which are clearly not observed. On the other hand, starting
from a $2p^{5}3d^{1}$ final-state configuration in a cubic crystal field,
does yield the correct intensities,\cite{Groot90d0} plus several small
peaks. Our \textit{ab initio} results shown in blue in the figure, confirm
this interpretation. Within the atomic $2p^{5}3d^{1}$ excitonic picture the
interpretation in terms of $t_{2g}$ and $e_{g}$ excitations of the $L_{3}$
and $L_{2}$ edge for the four peaks might still be a reasonable starting
point, but one should realize that there is a considerable mixing between
states due to Coulomb interactions.

The cluster eigenstates cannot be represented by single Slater
determinants. For correlated TM compounds, the spectral line-shape is
governed by multiplet effects leading to involved spectral functions, not
obviously related to the density of states.\cite{Fink85, Groot90} The
spectra are therefore generally used as fingerprints which contain unique
features resembling the local ground-state properties. The energy of the
final state is determined by local atomic-like physics. The intensity with
which each state can be reached depends, via the optical selection rules, on
the ground-state symmetry and the polarization of the light. This can lead
to large spectral changes for small changes in the ground-state.\cite%
{Fink85, Groot90, Groot08} For example, a $p$ electron can only be excited
into an $d_{x^{2}-y^{2}}$ orbital with $x$ or $y$ polarized light, but not
with $z$ polarized light. Changing the orbital occupation can therefore lead
to a strong polarization dependence which for certain multiplets can be as
strong as 100\%.\cite{Chen92} Due to the strong TM $2p$ spin-orbit coupling
the XAS spectra are also sensitive to the spin of the ground-state.\cite%
{Thole85, Goedkoop88}

Theoretically, as well as experimentally, one finds that the monopole part
of the $2p$-$3d$ Coulomb interaction is larger than that of the $3d$-$3d$
interaction. \cite{Cowan81, Bocquet96,Tanaka94} This leads to strongly bound
excitons at the TM $L_{2,3}$ edge and allows one to describe the spectra
using MLFT. Besides these excitonic states, also excitations into non-bound
states are possible.\cite{Laan86} Such excitations essentially probe the
conduction bands of the compound. For NiO both excitations are clearly
visible in the experimental spectra shown in the bottom panel of Fig.$\,$\ref%
{ExpL23XAS}: The excitonic bound states give rise to sharp excitations which
extend upwards from 852 eV; they are seen to agree very well with our MLFT
spectra. At 865 eV the experimental NiO spectra show an \emph{edge jump}
where the cross section for photon absorption increases discontinuously.
This is the onset of $L_{3}$ excitations into the conduction-band continuum
without formation of bound-states. These continuum excitations are of course
not reproduced with MLFT. The $L_{2}$ excitations into bound excitons give
rise to the sharp features starting at 870 eV and captured by our MLFT.
Around 856 eV there is a slight disagreement between the theoretical and
experimental Ni-$L_{3}$ edge spectra which might be due to the neglect of
the Ni $4s$ orbitals in our cluster basis set.

Looking at NiO, MnO and SrTiO$_{3},$ one may notice that our calculations
reproduce the low-energy parts of the spectra better than the high-energy
parts. The former are most excitonic and therefore best described by the
small basis set in the cluster. It may furthermore be noticed that not only
the edge-jumps are absent in the calculation, but also the interference
between the excitonic excitations of the $L_{2}$ edge and the continuum
states of the $L_{3}$ edge. These interference effects give rise to Fano
like line-shapes present in the experiment, but not in the theory. The
effect is relative small as the interference between $2p_{j=3/2}$ and $%
2p_{j=1/2}$ states is forbidden in many channels. There is, however, a
substantial mixing of core-states due to Coulomb interactions, which could
be the main reason for the interference effects between continuum and
excitonic states of the $L_{3}$ and $L_{2}$ edge.

Nevertheless, the agreement between MLFT and experiment is rather good in
Fig.$\,$\ref{ExpL23XAS} for all three TM oxides. This agreement is as good
as --or even better than-- that obtained for calculations in which all standard MLFT
parameters are optimized to give the best fit to experiment. \cite{Groot08,Uehara97,Csiszar05A,Csiszar05B,Alders98} There are many parameters
in such a calculation and finding the best fit is not trivial. The use of 
\textit{ab initio} values for an otherwise equivalent MLFT calculation can
therefore be of great help to interpret x-ray absorption spectra and thus
also in the understanding of elastic resonant x-ray diffraction (RXD)
spectra,\cite{Benckiser11} and the resonant energy dependence of resonant
inelastic x-ray scattering (RIXS).\cite{Haverkort10,Glawion11} For systems
with lower local symmetry, the number of parameters is even larger, and so
is the need for values determined \textit{ab initio}.

Compared with other \textit{ab initio} methods used for the calculation of
the $L_{2,3}$ edges of correlated transition-metal compounds\cite%
{Ogasawara01, Ikeno05, Ikeno06, Ikeno09, Miedema11, Ikeno11, Kruger04,
Kruger10, Laskowski10} the current method preforms well. For $\mathrm{d}^{0}$
compounds, i.e. band insulators like SrTiO$_{3},$ very powerful methods
based on multiple-scattering formalisms,\cite{Kruger04, Kruger10} or the
Bethe-Salpeter equations,\cite{Laskowski10} are available. For Mott-Hubbard
or charge-transfer insulators, \textit{ab initio} configuration-interaction
calculations of high quality have been performed for finite-sized clusters.%
\cite{Ogasawara01, Ikeno05, Ikeno06, Ikeno09, Miedema11, Ikeno11} Our MLFT
method has the advantage that its one-electron basis functions exactly
describe the relevant bands for the infinite crystal and at the same time
localize so well that one can afford to include correlations beyond the LDA
for merely the TM $d$ orbitals. This allows for very efficient, but still
accurate many-body calculations in the framework of the well-studied
multiplet ligand field theory.

\subsection{$2p$ core-level photoemission of Ni$_{0.03}$Mg$_{0.97}$O}

\begin{figure}[tbp]
\includegraphics[width=0.5\textwidth]{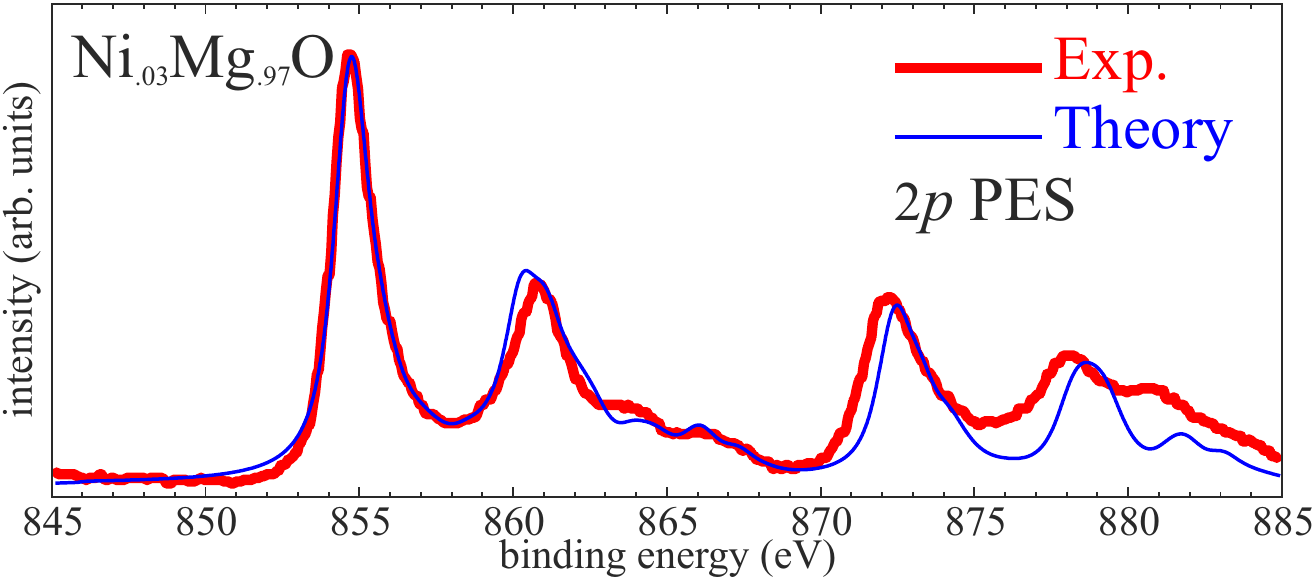}
\caption{(color online) Comparison of the experimental (thick red) and MLFT
(thin blue) Ni $2p$ core-level photoemission spectra of NiO in the impurity
limit. The experimental spectra are reproduced from Altieri \textit{et al.}.%
\protect\cite{Altieri00}}
\label{ExpNiO2pPES}
\end{figure}

Core-level photoemission is uninteresting from a one-electron point of view.
Core levels are atomic like, have no momentum-dependent dispersions, and
therefore delta-peaked densities of states. Accordingly, $2p$ photoemission
is expected to yield two spin-orbit-split peaks with intensity ratio $2:1$.
However, the experimental spectra\cite{Altieri00} from Ni $2p$ core-level
photoemission in Ni$_{0.03}$Mg$_{0.97}$O exhibit four distinct spectral
features, as shown in Fig.$\,$\ref{ExpNiO2pPES}. Emission from the Ni $%
2p_{3/2}$ level gives rise to the structure between 852 eV and 868 eV and
emission from the $2p_{1/2}$ level to the structure between 870 eV and 886
eV. The structure between 860 and 868 eV originates from multiplet
excitations with main character $\underline{2p}_{3/2}3d^{8}$ (with the
underbar indicating a hole) while the peak centered at 855 eV belongs to an
excitation with main final-state character $\underline{2p}_{3/2}3d^{9}%
\underline{L}.$ It is the strong Coulomb attraction between the $2p$ core
hole and the $3d$ electron which causes the latter state to have lower
energy than the former, and thus screens the core hole by driving charge in
from the ligand. Between 870 and 886 eV, this spectrum of screened and
un-screened states is repeated, but now for excitations from the $2p_{1/2}$
core level. {Our MLFT spectra agree well with the experimental spectra \cite%
{Altieri00} and with MLFT calculations for Ni impurities in MgO with fitted
parameters.\cite{Altieri00}} The resulting interpretations of the experiment
are the same.

These Ni $2p$ core-level photoemission spectra are strikingly different from
the previously considered Ni $2p$ x-ray absorption spectra in the bottom
panel of Fig.$\,$\ref{ExpL23XAS}.\cite{Laan86} In core-level x-ray
absorption, a TM core electron is excited into the $3d$ shell of the same
atom, whereby the sample remains locally neutral. The $3d$ electron can bind
with the core-hole left behind and thereby screen the core-hole potential.
This gives rise to the strong excitonic peaks seen in the x-ray absorption
spectrum. In core-level photoemission, a core electron is emitted from the
sample (excited into the vacuum) and can therefore not screen the core hole
left behind. The core-hole is either screened by the surrounding
ligands or left un-screened, which gives rise to higher-energy excitations.

Photoemission spectra are generally not excitonic. It might therefore seem
strange to use MLFT to calculate those spectra. Nevertheless, experience has
shown that many features of photoemission from correlated transition-metal
compounds can be captured by full multiplet theory for a local cluster.
Photoemission combined with cluster calculations has contributed greatly to
our understanding of correlated TM and RE compounds.\cite{Bocquet96,
Ghijsen88} The influence of non-local screening, i.e. the effect of the
material being a solid and not a single impurity, has been studied
experimentally by comparing the core-level photoemission from TM impurities
with that from the TM compounds.\cite{Altieri00} The main features of the Ni 
$2p$ photoemission spectra from Ni$_{0.03}$Mg$_{0.97}$O are the same as from
NiO. The largest bulk effect is a splitting of the peaks at $855$ and $873$
eV.

Important progress in understanding bulk valence photoemission from NiO has
been made recently by solving the LDA O $p$ Ni $d$ Wannier-orbital Hubbard
model in the dynamical mean-field approximation (LDA+DMFT)\cite{Kunes07A,
Kunes07B} and also by using the variational cluster approximation.\cite%
{Eder08} How important correlations between different Ni sites is, and
therefore how important the inclusion of dynamical non-local screening
effects are,\cite{Veenendaal93} remains an open question. On a different
level, MLFT is able to reproduce a substantial part of the photoemission
spectra, even though these are not excitonic.

\subsection{Non-resonant inelastic x-ray scattering at the \textit{M}$_{2,3}$
edge}

\begin{figure}[tbp]
\includegraphics[width=0.5\textwidth]{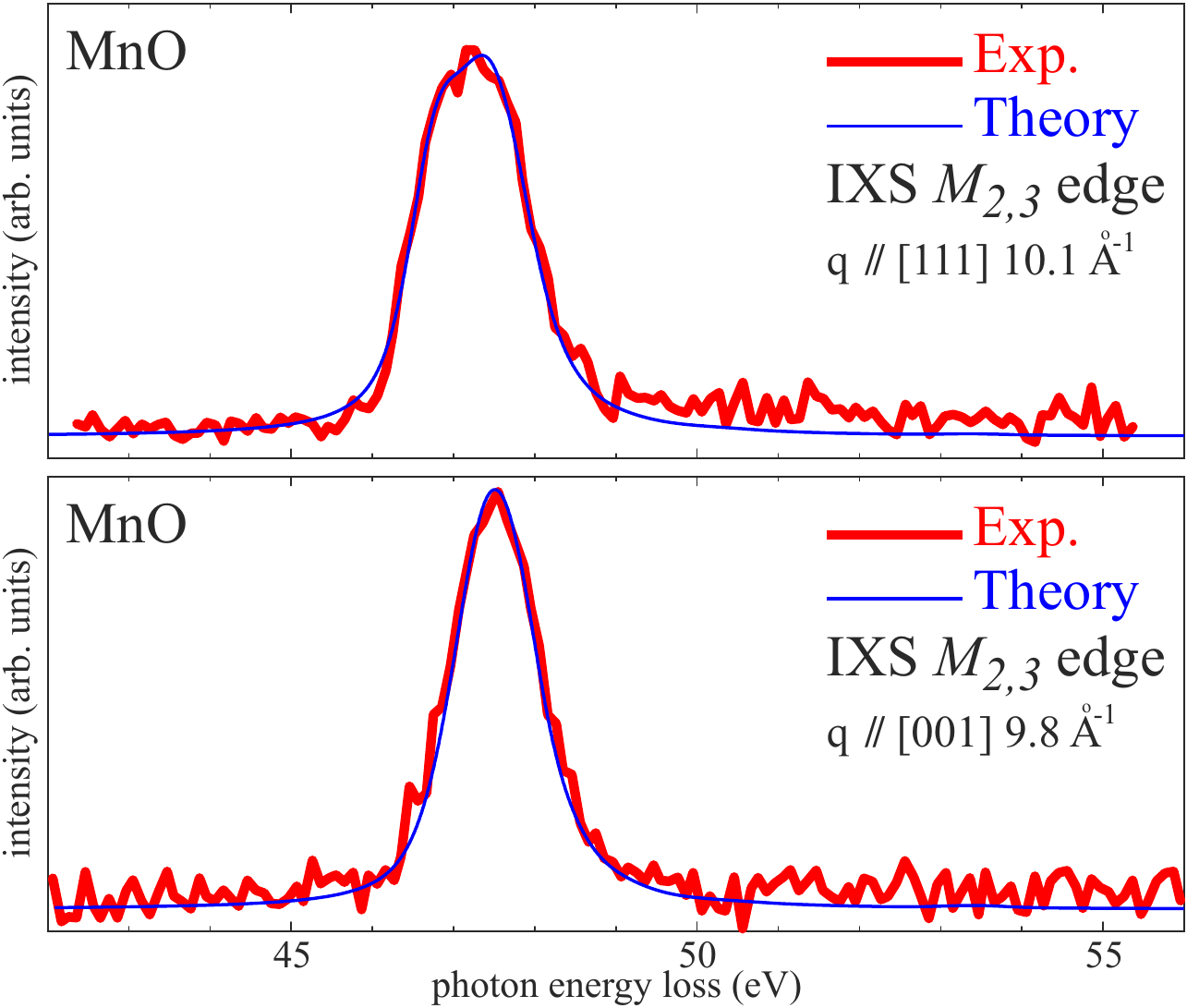}
\caption{(color online) Comparison of the experimental (thick red) and
theoretical (thin blue) Mn $3p$ ($M_{2,3}$ edge) non-resonant inelastic
x-ray scattering spectra of MnO at large momentum transfer ($\mathbf{q}$).
The two panels give spectra for different directions of momentum transfer.
They exhibit the generalized natural linear dichroism present for an
octupole transition in cubic symmetry. The experimental spectra are
reproduced from Gordon \textit{et al.}.\protect\cite{Gordon09}}
\label{ExpMnOM23IXS}
\end{figure}

In subsection A, we compared experiments and MLFT for core level x-ray
absorption (XAS). Now we shall discuss core-level spectra obtained with a
technique which from a theoretical point of view is very similar to x-ray
absorption, namely inelastic x-ray scattering (IXS). In XAS at the TM $%
M_{2,3}$ edge, a TM $3p$ core state is excited to a $3d$ conduction state by
absorption of a photon. The same excitation can be made when a photon is
scattered inelastically and only part of its energy is absorbed.\cite%
{Schulke07, Hamalainen02, Haverkort07, Gordon08} The major difference
between XAS and IXS is that, for the former, the energy of the photon has to
equal the absorption edge, whereas for the latter, the energy of the photon
should be (much) higher than the absorption edge since only a fraction of
its energy is absorbed. For XAS at the TM $M_{2,3}$ edge, the leading
interaction is of dipole character, i.e. one can use the long wave-length
limit. For IXS, the transferred momentum can be selected by looking at
different scattering angles and energies: for small momentum transfers,
dipole transitions are measured and for larger momentum transfers, octupole
transitions.\cite{Haverkort07} In Fig.$\,$\ref{ExpMnOM23IXS} we show the
non-resonant IXS at the $M_{2,3}$ edge in MnO for high magnitude momentum
transfers where octupole transitions are the strongest.

There is a clear difference between the $L_{2,3}$ ($2p$ to $3d$) and $M_{2,3}$ ($3p$ to $3d$) edges in MnO. One reason is that the spin-orbit
coupling constant for $3p$ is much smaller than for $2p$, \textit{e.g.} for
Mn, $\zeta _{2p}=6.85$ and $\zeta _{3p}=0.77$ eV. The splitting between the $M_{3}$ and $M_{2}$ edges is thus much smaller (not resolved in the
experimental spectra) than the splitting between the $L_{3}$ and $L_{2}$
edges. Another reason why the $L_{2,3}$ and the $M_{2,3}$ edges look different comes from the fact that the $3p$ wave-function is larger than the $2p$
wave-function due to the extra node. This leads to a smaller monopole part
of the Coulomb repulsion and larger multipole interactions between the $p$
and $3d$ orbitals. In general, for excitations within the same radial shell
the multiplet splittings are larger than the excitonic binding energy.\cite%
{Gupta11} This has important consequences. The highest-energy multiplets of
the $M_{2,3}$ excitations are pushed above the continuum edge and form broad
resonances instead of sharp mulitplets. The low-energy multiplets, on the
other-hand, are still sharp excitonic states. Due to the strict selection
rules applicable to XAS and IXS, one can choose the experimental geometry
such that only particular excitations are allowed. The spectra shown in Fig.$%
\,$\ref{ExpMnOM23IXS} are octupole dominated and only sensitive to the
low-energy excitonic features in the spectra. One would not be able to
reproduce the broad dipole resonances with MLFT.

One of the beauties of octupole transitions is that they show dichroism in
cubic symmetry.\cite{Gordon09} This can be seen in the two different panels
of Fig.$\,$\ref{ExpMnOM23IXS}, and is well reproduced by our theory. For a
dipole transition one can not distinguish cubic from spherical symmetry. (A
transition of angular momentum $L=1$ branches to a single irreducible
representation ($T_{1u}$) in cubic symmetry.) An octupole transition,
however, shows nice dichroism in cubic symmetry, whereby the momentum
transfer $\mathbf{q}$ for IXS takes the place of the light polarization $%
\mathbf{\epsilon }$ in XAS. A transition of angular momentum $L=3$ branches
to three different irreducible representations in cubic symmetry, namely $%
T_{1u}$, $T_{2u}$ and $A_{2u}$. As a consequence, the dichroic spectra can
be used to obtain detailed information about the differences in bonding of $%
t_{2g}$ and $e_{g}$ electrons. The shift in the spectral energy and the
change in spectral weight for excitations with $\mathbf{q}$ either in the $%
[111]$ (top panel) or $[001]$ (bottom panel) direction is related to the
different energy of the $t_{2g}$ and $e_{g}$ electrons and the difference in
occupation of these orbitals due to covalent bonding.

\subsection{Non-resonant inelastic x-ray scattering of \textit{d}-\textit{d}
excitations}

\begin{figure}[tbp]
\includegraphics[width=0.5\textwidth]{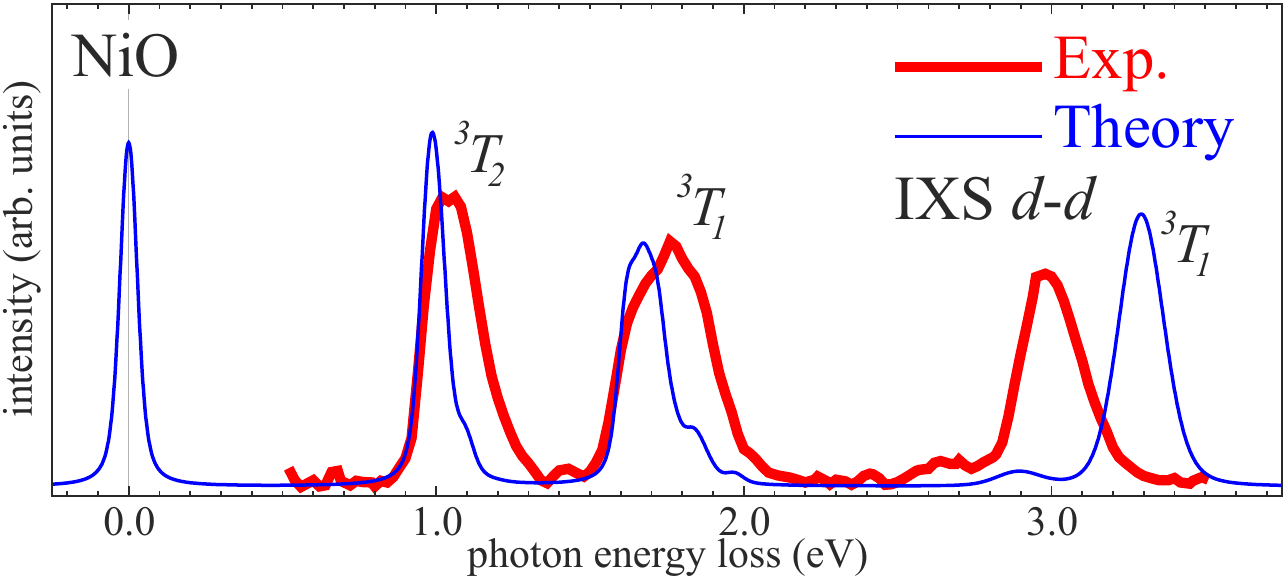}
\caption{(color online) Comparison of the experimental (thick red) and MLFT
(thin blue) non-resonant inelastic x-ray scattering intensity of low energy $%
d$-$d$ excitations. The experimental spectra are reproduced from Verbeni 
\textit{et al.}.\protect\cite{Verbeni09} }
\label{ExpNiOddIXS}
\end{figure}

This last section in which we compare MLFT with experiment deals with
low-energy excitations without a core hole. It has recently been shown that
surprisingly strong $d$-$d$ or crystal-field excitations can be observed in
NiO with non-resonant IXS for large momentum transfers. \cite%
{Larson07,Verbeni09, Hiraoka11} These spectra contain similar information as
the weak $d$-$d$ excitations inside the optical gap observed with optical
spectroscopy.\cite{Newman59, Powell70} The difference between IXS and optics
is that with optics these transitions, being even in parity, are forbidden
and only become allowed by simultaneous excitation of a phonon and a
crystal-field excitation. This makes a quantitative interpretation of
optical $d$-$d$ excitations involved. The interpretation of the non-resonant
IXS is, on the other hand, straight forward and allows for a quantitative
comparison.\cite{Haverkort07, Veenendaal08}

In Fig.$\,$\ref{ExpNiOddIXS} we show the experimental\cite{Verbeni09} and
theoretical non-resonant IXS spectra for a powder of NiO at large momentum
transfer (averaged over a transfer of $7.3-8.0$ {\AA } $^{-1}$). These
spectra are governed by quadrupole and hexadecapole transitions between the $%
3d$ orbitals. The non-resonant IXS excitations are spin-conserving. Locally
the Ni ground-state configuration is $\mathrm{d}^{8}$ with the $t_{2g}$
orbitals fully occupied and the $e_{g}$ orbitals half filled with $\langle
S^{2}\rangle =2$, i.e. $S=1$. In the one-electron picture, one can make a
single excitation going from the $t_{2g}$ shell to the $e_{g}$ shell, which
has an experimental energy of about 1.1 eV. This is the peak of $T_{2g}$
final state symmetry in the experiment. In principle one could also excite
two $t_{2g}$ electrons simultaneously into the $e_{g}$ sub-shell. This would
give rise to a single peak at twice the energy. In a pure one-electron
picture the double excitation is forbidden because non-resonant IXS couples
a single photon to a single electron. Using full multiplet theory, however,
both excitations have a finite intensity. This has to do with the strong $%
t_{2g}$-$e_{g}$ multiplet interaction which mixes, for the excited states,
the single Slater determinants. One even finds three peaks instead of two.
The first peak indeed corresponds to an excitation of a single $t_{2g}$
electron into the $e_{g}$ sub-shell. The second peak is roughly the
simultaneous excitation of two $t_{2g}$ electrons into the $e_{g}$ subshell.
Finally, in order to understand the third peak, one should realize that the $%
t_{2g}\left( xy\right) $ electron is Coulomb repelled more from an $%
e_{g}\left( x^{2}-y^{2}\right) $ electron than from an $e_{g}\left(
3z^{2}-1\right) $ electron because of the larger overlap of densities. This
leads to multiplet splitting of the $t_{2g}^{5}e_{g}^{3}$ states and to
mixing of $t_{2g}^{5}e_{g}^{3}$ and $t_{2g}^{4}e_{g}^{4}$ states.

One could also have understood the energy and number of excitations by
starting from spherical symmetry where Coulomb repulsion splits the $S=1$
states into a lowest state of $^{3}F$ symmetry and an excited state of $%
^{3}P $ symmetry. In cubic symmetry the $^{3}F$ states branch into a $%
^{3}A_{2}$ ground-state, a $^{3}T_{2}$ first excited state, and a $^{3}T_{1}$
second excited state. The $^{3}P$ state branches to a state of $^{3}T_{1}$
symmetry, which can mix with the highest excited state branching from the $%
^{3}F$ state. Such multiplet effects are hard to capture at the DFT level.
Recent time dependent DFT calculations with the LDA+$U$ functional do show
Frenkel excitons ($d$-$d$ excitations) within the optical gap, but they
cannot reproduce the correct number of multiplet states.\cite{Lee10}

Let us finally have a closer look at the comparison between the experimental
and MLFT crystal-field excitations in NiO. MLFT gets the lowest excitation $%
\left( ^{3}T_{2}\right) $ 5\% too low and the highest $\left(
^{3}T_{1}\right) $ 10\% too high. As the $^{3}T_{2}$ energy is mainly
determined by one-electron interactions, we conclude that the $e_{g}-t_{2g}$
splitting due to covalency in our LDA based calculation is 5\%
underestimated. At the same time, the multiplet splitting due to the Coulomb
repulsion, i.e. the values of the $F_{dd}^{\left( 2\right) }$ and $%
F_{dd}^{\left( 4\right) }$ Slater integrals, are 10\% overestimated. The
later could be a result of neglecting the screening of the multipole
interactions, but not necessarily, because there are additional channels in
which two $3d$ electrons can scatter into two higher excited states due to
Coulomb repulsion. This gives rise to a multiplet-dependent screening, not
easily described with a single screening parameter.\cite{Gupta11}

\section{Conclusions}

We have shown how multiplet ligand-field theory (MLFT) calculations can be
based on \textit{ab-initio} LDA solid-state calculations, in a similar way
as originally devised by Gunnarson \textit{et al.}\cite{Gunnarsson89, Anisimov91} and recently done for LDA+DMFT calculations. The
resulting method could be named LDA+MLFT. The theory is very well suited for
the calculation of local ground-state properties and excitonic spectra of
correlated transition-metal and rare-earth compounds. Our TM $d$ Wannier
orbitals, which \emph{together} with the O $p$ Wannier orbitals span the LDA
TM $d$- \emph{and} O $p$-bands, are quite similar to atomic orbitals, and
this justifies many previous studies using the latter.

We compared several experimental spectra (XAS, non-resonant IXS, PES) for
SrTiO$_{3}$, MnO$,$ and NiO with our \emph{ab initio} multiplet ligand-field
theory and found overall satisfactory agreement, indicating that our
ligand-field parameters are correct to better than 10\%. The covalency seems
to be slightly underestimated and the Slater integrals for the higher
multipole interactions overestimated. The method is expected to provide
insights to the local properties of transition-metal compounds with only
modest computational efforts.

We would like to thank Eva Pavarini, Ove Jepsen, and Olle Gunnarsson for
fruitful discussions. Support by the Deutsche Forschergemeinschaft through
FOR 1346 is gratefully acknowledged.

\appendix

\section{Definition of O and TM orbitals, covalency and formal valence}
\label{appCovalence}

\begin{figure}[h]
\includegraphics[width=0.5\textwidth]{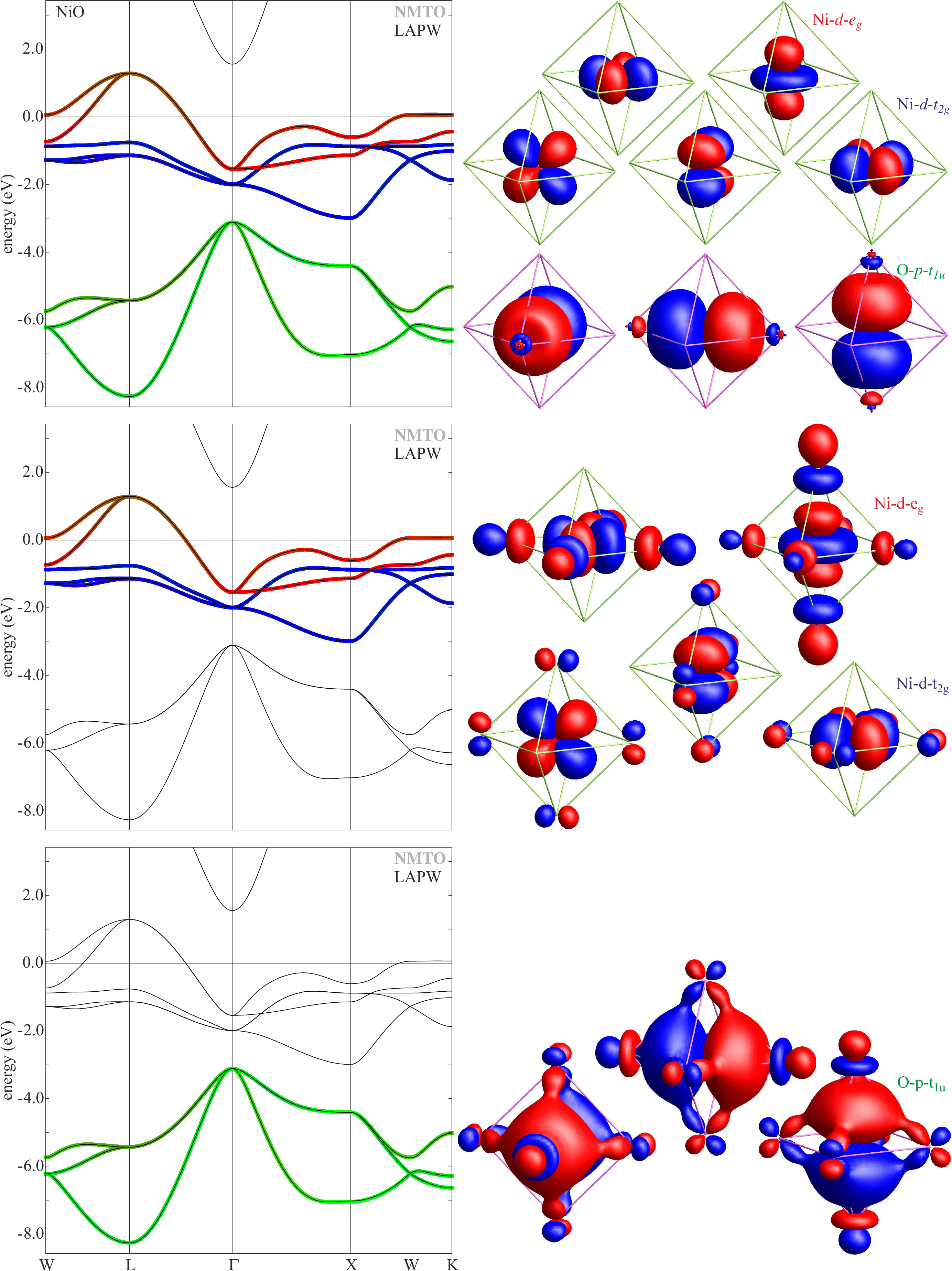}
\caption{(color online) NiO band structure (left) and Wannier orbitals (right) for three different basis sets. Top panels: Including both the Ni $d$ and O $p$ orbitals. Middle panels: Including only the Ni $\mathrm{d}$ orbitals. Bottom panels: Including only the O $\mathrm{p}$ orbitals.}
\label{Fig_orbital_definitions}
\end{figure}

It is a general praxis to talk about Ni-$d$ and O-$p$ orbitals, even in a solid. The definition of local orbitals in a solid is not always clear. In this paper we choose such orbitals as generalized Wannier orbitals of a given local symmetry. For our materials, i.e. transition metal oxides, there are two useful definitions for the Wannier orbitals, depending on the size of the basis set used and the energy bands they span. These two different definitions are often used in an ambivalent way. Here we explain the definition of the two different set of orbitals, by the example of NiO. 

In Fig. \ref{Fig_orbital_definitions} we show in the top row the Ni $d$ and O $p$ Wannier orbitals which are most atomic like. Linear combinations of these 8 orbitals span the 8 bands shown on the left of the top row in Fig. \ref{Fig_orbital_definitions}. Due to covalency, the Ni $d$ orbitals defined in this way are occupied by more than 8 electrons. At the same time, the O $p$ Wannier orbitals are occupied by less than 6 electrons, i.e. they have holes. When forming band states, the Ni $d$ and O $p$ orbitals mix and part of the O $p$ character ends above the Fermi energy. One should not think of these extra electrons or holes as mobile charge carriers. These partial occupations are just a result of the choice of the basis set used, which is different from the eigenbasis, the band states.

A different choice of the Wannier orbitals, closer to the eigenfunctions, can be seen in the lower two rows of Fig. \ref{Fig_orbital_definitions}. Here we show from top to bottom the Ni $\mathrm{d}$ ($\mathrm{e_g}$ and $\mathrm{t_{2g}}$) and O $\mathrm{p}$ $\mathrm{t_{1u}}$ orbitals. The Ni $\mathrm{d}$ orbitals are those 5 orbitals which together span the Ni $\mathrm{d}$ bands in the energy range from -3 to 2 eV. The O $\mathrm{p}$ orbitals are those 3 orbitals which together span the O $\mathrm{p}$ bands in the energy range from -8 to -3 eV.

In order to distinguish the two different sets of Wannier orbitals we have throughout this paper used \textit{italic} font for the more localized atomic like orbitals and $\mathrm{roman}$ font for the more extended orbitals. The difference in notation is quite subtle, but in almost all cases one can understand from the context which definition is meant.

The two basis sets of either atomic like Ni $d$ and O $p$ orbitals or more delocalized Ni $\mathrm{d}$ and O $\mathrm{p}$ orbitals span the same bands and can thus be expressed in terms of linear combinations of each other. The unitary transformation between the two sets of Wannier orbitals is such that it diagonalizes the covalent interaction between the Ni $d$ and O $p$ orbitals. The Ni $d$ and O $p$ orbitals interact, whereas the Ni $\mathrm{d}$ and O $\mathrm{p}$ orbitals are non interacting at the one particle or LDA level. The 5 Ni $\mathrm{d}$ orbitals span the 5 $\mathrm{d}$ bands exactly and the 3 O $\mathrm{p}$ orbitals spand the 3 O $\mathrm{p}$ bands exactly. The O $\mathrm{p}$ orbitals are bonding combinations of the TM $d$ and O $p$ orbitals. The TM $\mathrm{d}$ orbitals are anti bonding combinations of the TM $d$ and O $p$ orbitals.

With the use of the Ni $\mathrm{d}$ (O $\mathrm{p}$) Wannier orbitals, which span only the Ni $\mathrm{d}$ (O $\mathrm{p}$) bands one can define the formal valence of Ni in NiO. It is common to state that O is $2-$, i.e. has an occupation of $\mathrm{p}^{\mathrm{6}}$ and Ni is $2+$, i.e. has an occupation of $\mathrm{d}^{\mathrm{8}}$. If one counts the electrons in the Wannier orbitals that separately span the Ni or O bands one immediately reproduces the formal valence. The occupation numbers are different if one looks at the more atomic like Ni $d$ and O $p$ orbitals. For these orbitals covalence introduces holes in the O Wannier orbitals and extra electrons in the Ni Wannier orbitals. For the oxides described in the present paper, the occupations of the two kinds of Wannier orbitals are: NiO ${3}\mathrm{d}^{8}\approx 2p^{5.4}3d^{8.6}$, MnO $3\mathrm{d}^{5}\approx 2p^{5.5}3d^{5.5}$, and SrTiO$_{3}$ $3\mathrm{d}^{0}\approx \left(2p^{5.7}\right)_{3} 3d^{0.9}$.

Let us note that for our purposes, the
Wannier $\mathrm{d}$-orbitals are \emph{not} sufficiently localized.
Nevertheless, in early LDA+DMFT calculations which could handle only a few
correlated orbitals, even more "downfolded" $\mathrm{t}_{2g}$ or $\mathrm{e}%
_{g}\,$Wannier-orbitals$\,${\cite{Pavarini05, Yamasaki06}} were used by
necessity; they clearly exhibit the covalencies.\cite{Zurek05} 

\section{Computational details}
\label{appComp}

\begin{table*}[tbp]
\begin{tabular}{l|r@{.}lr@{.}lr@{.}lr@{.}lr@{.}lr@{.}lr@{.}lr@{.}lr@{.}lr@{.}lr@{.}lr@{.}lr@{.}lr@{.}lr@{.}lr@{.}l}
\hline\hline
\rule{0pt}{4ex} & \multicolumn{2}{c}{$V_{e_{g}}$} & \multicolumn{2}{c}{$%
V_{t_{2g}}$} & \multicolumn{2}{c}{10Dq} & \multicolumn{2}{c}{$T_{pp}$} & 
\multicolumn{2}{c}{$\zeta_{3d}$} & \multicolumn{2}{c}{$F^{(2)}_{dd}$} & 
\multicolumn{2}{c}{$F^{(4)}_{dd}$} & \multicolumn{2}{c}{$\zeta_{2p}$} & 
\multicolumn{2}{c}{$F^{(2)}_{2p3d}$} & \multicolumn{2}{c}{$G^{(1)}_{2p3d}$}
& \multicolumn{2}{c}{$G^{(3)}_{2p3d}$} & \multicolumn{2}{c}{$\zeta_{3p}$} & 
\multicolumn{2}{c}{$F^{(2)}_{3p3d}$} & \multicolumn{2}{c}{$G^{(1)}_{3p3d}$}
& \multicolumn{2}{c}{$G^{(3)}_{3p3d}$} & \multicolumn{2}{c}{} \\ \hline
NiO         & 2&06 & 1&21 & 0&56 & 0&72 & 0&08 & 11&14 & 6&87 & 11&51 & 6&67 & 4&92 & 2&80 & 1&40 & 12&87 & 15&89 & 9&58 & \multicolumn{2}{c}{}\\ 
MnO         & 1&92 & 1&15 & 0&67 & 0&53 & 0&04 &  9&35 & 5&78 &  6&85 & 5&29 & 3&77 & 2&14 & 0&77 & 10&93 & 13&56 & 8&15 & \multicolumn{2}{c}{}\\ 
SrTiO$_{3}$ & 4&03 & 2&35 & 1&79 & 0&99 & 0&02 &  8&38 & 5&25 &  3&78 & 4&23 & 2&81 & 1&59 & 0&43 &  9&85 & 12&08 & 7&35 & \multicolumn{2}{c}{}\\
\hline\hline
\rule{0pt}{4ex} & \multicolumn{4}{c|}{$\big[00\frac{1}{2}\big]$} & 
\multicolumn{20}{c|}{$\big[\frac{1}{2}\frac{1}{2}0\big]$} & 
\multicolumn{6}{c}{$\big[\frac{1}{2}\frac{1}{2}\frac{1}{2}\big]$} & \multicolumn{2}{c}{} \\ 
& 
\multicolumn{2}{c}{$d_{z^2}p_{z}$} & \multicolumn{2}{c|}{$d_{xz}p_{x}$} & 
\multicolumn{2}{c}{$p_xp_y$} & \multicolumn{2}{c}{$p_yp_x$} & 
\multicolumn{2}{c}{$p_xp_x$} & \multicolumn{2}{c}{$p_zp_z$} & 
\multicolumn{2}{c}{$d_{z^2}d_{z^2}$} & \multicolumn{2}{c}{$d_{\mathsf{\scriptscriptstyle{XY}}}d_{\mathsf{\scriptscriptstyle{XY}}}$} & 
\multicolumn{2}{c}{$d_{xy}d_{xy}$} & \multicolumn{2}{c}{$d_{xz}d_{xz}$} & 
\multicolumn{2}{c}{$d_{xz}d_{yz}$} & \multicolumn{2}{c|}{$d_{xy}d_{z^2}$} & 
\multicolumn{2}{c}{$d_{xy}p_{y}$} & \multicolumn{2}{c}{$d_{xy}p_{z}$} & 
\multicolumn{2}{c}{$d_{z^2}p_{z}$} & \multicolumn{2}{c}{} \\ \hline
NiO         & \textbf{ 1}&\textbf{19} & \textbf{-0}&\multicolumn{1}{l|}{\textbf{60}} & \textbf{0}&\textbf{38} & \textbf{0}&\textbf{38} & \textbf{0}&\textbf{25} & 
-0&10 & -0&01 & -0&08 & -0&20 &  0&06 &  0&04 &  0&\multicolumn{1}{l|}{04} & -0&03 & -0&02 &  0&00 & \multicolumn{2}{c}{} \\ 
MnO         & \textbf{ 1}&\textbf{11} & \textbf{-0}&\multicolumn{1}{l|}{\textbf{57}} & \textbf{0}&\textbf{28} & \textbf{0}&\textbf{28} & \textbf{0}&\textbf{19} & 
-0&09 & -0&04 & -0&06 & -0&26 &  0&08 &  0&05 &  0&\multicolumn{1}{l|}{04} & -0&03 & -0&02 &  0&02 & \multicolumn{2}{c}{} \\ 
SrTiO$_{3}$ & \textbf{ 2}&\textbf{33} & \textbf{-1}&\multicolumn{1}{l|}{\textbf{18}} & \textbf{0}&\textbf{42} & \textbf{0}&\textbf{34} & \textbf{0}&\textbf{24} & 
-0&07 & --&---& --&---& --&---& --&---& --&---& --&\multicolumn{1}{l|}{---}& --&---& --&---& --&--- & \multicolumn{2}{c}{} \\
 \hline\hline
\rule{0pt}{4ex} & \multicolumn{14}{c|}{$\big[001\big]$} & 
\multicolumn{10}{c|}{$\big[0\frac{1}{2}1\big]$} & \multicolumn{4}{c|}{$\big[11\frac{1}{2}\big]$} & \multicolumn{4}{c}{$\big[000\big]$}\\ 
& 
\multicolumn{2}{c}{$p_zp_z$} & \multicolumn{2}{c}{$p_xp_x$} & \multicolumn{2}{c}{$p_zp_z$} &
\multicolumn{2}{c}{$p_xp_x$} & \multicolumn{2}{c}{$p_yp_y$} & 
\multicolumn{2}{c}{$d_{z^2}d_{z^2}$} &  
\multicolumn{2}{c|}{$d_{xz}d_{xz}$} & 
\multicolumn{2}{c}{$d_{xy}p_{x}$} & \multicolumn{2}{c}{$d_{yz}p_{z}$} & \multicolumn{2}{c}{$d_{z^2}p_{y}$} & \multicolumn{2}{c}{$d_{z^2}p_{z}$} & \multicolumn{2}{c|}{$d_{xz}p_{x}$} &
\multicolumn{2}{c}{$d_{xz}p_{y}$} & \multicolumn{2}{c|}{$d_{xy}p_{x}$} & 
\multicolumn{2}{c}{$\epsilon_p$} & \multicolumn{2}{c}{$\epsilon_d$} \\ \hline
NiO         &  0&02 & -0&04 & --&---& --&---& --&---& -0&01 & -0&\multicolumn{1}{l|}{03} &  0&00 & 0&02 & -0&03 &  0&00 &  0&\multicolumn{1}{l|}{00} &  0&00 &  0&\multicolumn{1}{l|}{00} & -4&75 &-1&35 \\ 
MnO         &  0&05 & -0&03 & --&---& --&---& --&---& -0&06 & -0&\multicolumn{1}{l|}{05} &  0&00 & 0&02 & -0&03 & -0&01 &  0&\multicolumn{1}{l|}{00} &  0&00 &  0&\multicolumn{1}{l|}{00} & -5&22 &-0&39  \\ 
SrTiO$_{3}$ & -0&01 & -0&11 &  0&06 & -0&02 & -0&02 &  0&05 & -0&\multicolumn{1}{l|}{13} & -0&02 & 0&05 & -0&03 &  0&00 & -0&\multicolumn{1}{l|}{02} & -0&01 & -0&\multicolumn{1}{l|}{02} & -1&53 & 3&31  \\ 
\hline\hline
\end{tabular}%
\caption{Upper panel: MLFT parameters obtained
from LDA. The Slater integrals are obtained from the spherical averaged
Wannier orbitals. Lower two panels: One-electron tight binding parameters as obtained from the LDA TM-$d$ O-$p$ Wannier-orbital set. The hopping is from the first to the second orbital displaced by the vector [\textit{abc}]. [000] denote on-site energies. Shown are only those values larger than 10 meV. The bold numbers enter in
the MLFT calculations, the normal font longer range
hopping integrals are truncated in the cluster approximation. For
the $p$-$p$ hopping in the $[001]$ direction of SrTiO$_3$ the
first two values listed concern hopping along an O-Ti-O bond. The last three
values concerns O-O hopping in the Sr-O plane. The
notation for the $e_g$ orbitals is such that $d_{z^2} \equiv d_{3z^2-1}$ and $d_{\mathsf{\scriptscriptstyle{XY}}} \equiv d_{x^2-y^2}$. All values in electron volt.}
\label{ParameterTable}
\end{table*}

The selfconsistent LDA\cite{Ceperley80} LAPW calculations were performed
with the \textsc{Wien2k} code\cite{Blaha90} using a plane-wave cut off of $%
k_{max}\times R_{MT}=8$, with $R_{MT}$ the smallest MT-sphere radius and $%
k_{max}$ the largest $k$-vector. The $N$MTO calculations were done with the
Stuttgart code\cite{Stuttgart}, had $N\mathrm{=}1,$ and all partial waves
downfolded, except TM $d$ and O $p$.\cite{Andersen00,Andersen01,Andersen03,
Zurek05, Pavarini05} The LAPW warped potential (spherical inside the LAPW MT
spheres) was least-squares fitted to an overlapping MT potential (OMT) with
the recently developed OMTA code\cite{Zwierzycki09} and was used in the $N$%
MTO calculations. The radii of the hard screening spheres were 70\% of the
OMT radii.

The material-dependent settings are as follows:

NiO. Space-group Fm-3m (225) $a$=4.177$\,${\AA }, Ni at Wyckoff position 4a
and O at 4b. MT radii for Ni 2.08$a_{0},$ and for O 1.84$a_{0}$. $a_{0}$=0.5292 \AA, is the Bohr radius. OMT radii
for Ni 2.2$a_{0},$ for O 2.5$a_{0},$ and for an additional empty sphere at
Wyckoff position 8c 1.6$a_{0}$. The expansion energies were $-5.2$ and $-1.2$
eV.

MnO. Space-group Fm-3m (225) $a$=4.4248$\,${\AA }, Mn at Wyckoff position
4a, O at 4b. MT radii for Mn 2.20$a_{0},$ and for O 1.95$a_{0}$. OMT radii
for Mn 2.3$a_{0},$ for O 2.7$a_{0},$ and for an additional empty sphere at
Wyckoff position 8c 1.7$a_{0}.$ Expansion energies $-5.0$ and $-1.0$ eV.

SrTiO$_{3}.$ Space-group Pm-3m (221) $a$=3.905$\,${\AA }, Ti at Wyckoff
position 1a, Sr at position 1b, and O at position 3d. MT radii for Ti 2.32$%
a_{0}$, for Sr 2.00$a_{0}$, and for O 1.36$a_{0}$. OMT radii for Ti 2.4$%
a_{0},$ for Sr 3.8$a_{0}$ and for O 2.0$a_{0}.$ Expansion energies $-2.6$
and $+1.5$ eV.

The $N$MTO band-structures and densities of states as presented in Fig.s \ref%
{NiOLDA} and \ref{LDACompareNiOMnOSrTiO3} were calculated from the
real-space TM $3d$ O $2p$ Wannier-orbital (tight-binding) representation of
the LDA Hamiltonian, neglecting hops between sites more distant than 2.5$a$.
The tight-binding parameters larger than 10 meV are presented in the
lower panels of Table \ref{ParameterTable}. Only the nearest-neighbor hopping
integrals (bold faced) enter in the cluster calculations. In SrTiO$_{3},$
the point symmetry of O is merely tetragonal so that the $p$ orbital
pointing towards the Ti atom is slightly different from those pointing
perpendicular to the Ti-O bond (and \emph{e.g.} towards Sr). {For a
discussion of the bonding between the O $p$ and Sr $d$ orbitals and how this
changes the different O $p$ Wannier functions see Pavarini \textit{et al.}%
\cite{Pavarini05}.} Due to the two different types of O $p$ Wannier
functions, the relation between the hopping integrals and the cubic
ligand-field parameters is slightly more involved than those valid for O in
cubic symmetry and given in Sect. I. In general (for all symmetries), the
ligand field parameters can be found by block tri-diagonalization of the
tight-binding Hamiltonian of the cluster with respect to the TM $d$ orbitals. For more details see Appendix \ref{appBasisSize}

Wannier orbitals have tails on the neighboring sites, although most of the
orbital weight is close to the nucleus at its center (Fig.$\,$\ref{NiORofr}). The tails lead to long-ranged hopping integrals and their values are
given in the lowest panels of Table \ref{ParameterTable}. It should be
noticed that in order for MLFT to work properly with a basis set of LDA
based Wannier orbitals, it is important to have TM $3d$ atomic-like
character for $r\lesssim {1.5}\,$\AA, but it is not essential to have
hopping limited to the first-nearest neighbors.

The multipolar part of the Coulomb integrals is calculated by directly
integrating the wave-functions. In order to obtain numerically stable
integrals, the Wannier orbitals were expanded in radial wave-functions times
spherical harmonics, an approximation for which a set of Slater integrals
can be introduced. Sufficiently accurate results are obtained when different
radial wave-functions for the $t_{2g}$ and $e_{g}$ orbitals are used. The
core wave-functions, are calculated using the Hartree-Fock method.\cite%
{Cowan81} For reasons of space, Table \ref{ParameterTable} only gives the
Slater integrals for the radial functions averaged over $t_{2g}$ and $e_{g};$
the difference between these integrals for NiO are given as an inset in Fig.$%
\,$\ref{NiORofr}. The spin-orbit coupling constants have been calculated
using a spherical approximation, including only the local $d$ character at
the TM site. The resulting constant is the same for the $t_{2g}$ and $e_{g}$
orbitals.

\begin{table}[tbp]
\begin{tabular}{l|r@{.}lr@{.}lr@{.}lr@{.}l}
\hline\hline
& \multicolumn{2}{c}{$U_{3d,3d}$} & \multicolumn{2}{c}{$\Delta$} & 
\multicolumn{2}{c}{$U_{2p,3d}$} & \multicolumn{2}{c}{$U_{3p,3d}$} \\ \hline
NiO & 7 & 3 & 4 & 7 & 8 & 5 & -- & --- \\ 
MnO & \phantom{0}5 & 5\phantom{00} & \phantom{0}8 & 0\phantom{00} & %
\phantom{0}7 & 2\phantom{00} & \phantom{0}5 & 5\phantom{00} \\ 
SrTiO$_{3}$ & 6 & 0 & 6 & 0 & 8 & 0 & -- & --- \\ \hline\hline
\end{tabular}%
\caption{Multiplet ligand field theory parameters taken from experiment.
Note that the experiments shown in this paper are not very sensitive to
these parameters. For a more thoroughly discussion on these parameters we
refer to the papers by Bocquet (Ref. \onlinecite{Bocquet96}) or Tanaka (Ref. 
\onlinecite{Tanaka94}). All values in electron volt.}
\label{ParameterTableExperiment}
\end{table}

The parameters fitted to experiments are shown in Table \ref%
{ParameterTableExperiment}. These values are in good agreement with those in
the literature.\cite{Bocquet96, Tanaka94} One should realize that since
x-ray absorption involves a charge neutral excitation, it is not very
sensitive to $U$ and $\Delta $. The experiments discussed in this paper were
chosen to be most sensitive to the calculated values (Table \ref%
{ParameterTable}).

The $N$MTO method\cite{Andersen00, Andersen01, Andersen03} constructs the
basis set of Ni $d$ plus O $p$ localized orbitals by first constructing such
a set for \emph{each} of the $N+1$ energies, $\epsilon _{0},...,\epsilon
_{N},$ chosen to span the energy range of interest. In such a set of
zeroth-order $\left( N\mathrm{=}0\right) $ MTOs, each of the orbitals is a
solution of Schr\"{o}dinger's differential equation for the overlapping MT
potential for the chosen energy, but has \emph{kinks} at all Ni and O (but
not \textit{e.g.} Sr) hard spheres.\cite{kinkcancellation} Those hard spheres are
chosen to be a bit smaller than touching and not to coincide with a node of
the radial wave-function. The Ni $d_{xy}$ 0MTO, for instance, is now defined
by the hard-sphere boundary condition that all its $p$-projections on all O
spheres and all its $d$-projections on all Ni spheres, \emph{except }$d_{xy}$
on the \emph{own} Ni sphere, \emph{vanish.} And equivalently for the other
members of the 0MTO basis set. This hard-sphere boundary condition is what
localizes the 0MTOs, unless there are wave-functions at the chosen energy
with main characters different from those of the 0MTOs in the basis. The
condition that each 0MTO solves Schr\"{o}dinger's equation, except for kinks
in the Ni $d$ and O $p$ channels, means that each 0MTO is smooth in all 
\emph{other} channels. This is accomplished by constructing that set of
wave-equation solutions in the hard-sphere interstitial, the so-called
screened spherical waves, whose phase shifts are the hard-sphere ones for
all Ni $d$ and O $p$ channels, except the eigen-channel, and has the proper
phase shifts for all other channels (such as Ni $s$ and Sr $d$). The
screened spherical waves then gets augmented inside the overlapping MT
spheres to become the 0MTOs. Finally, the $N+1$ different 0MTO basis sets
are contracted into one, the $N$MTO set, which spans the solutions of Schr%
\"{o}dinger's equation at all $N+1$ energies. The $N$MTOs have
discontinuities in merely the $\left( 2N+1\right) $th derivatives at the
hard spheres and are therefore smooth if $N>0$. The contraction ($N$%
-ization) delocalizes the $N$MTO to a degree which depends on how much the
neighboring 0MTOs vary over the $N+1$ energies. This is so, because for an
energy-independent set of orbitals, the energy dependence of a radial Schr%
\"{o}dinger-equation solution must be provided by the tails of the
neighboring orbitals.\cite{Andersen70} The delocalization is further
enhanced by symmetrical orthonormalization of the $N$MTOs into Wannier
orbitals, and this depends on the overlap between neighboring $N$MTOs.
Nevertheless, as seen in Fig.$\,$\ref{NiOLDA}, our Ni $d$ plus O $p$ Wannier
orbitals are as localized as can be expected, and --in fact-- much better
than Wannier orbitals derived from a large set of energy-\emph{in}dependent
orbitals.\cite{07Solovyev}

\section{Double-counting correction}
\label{appDouble}

\begin{figure}[h]
\includegraphics[width=0.5\textwidth]{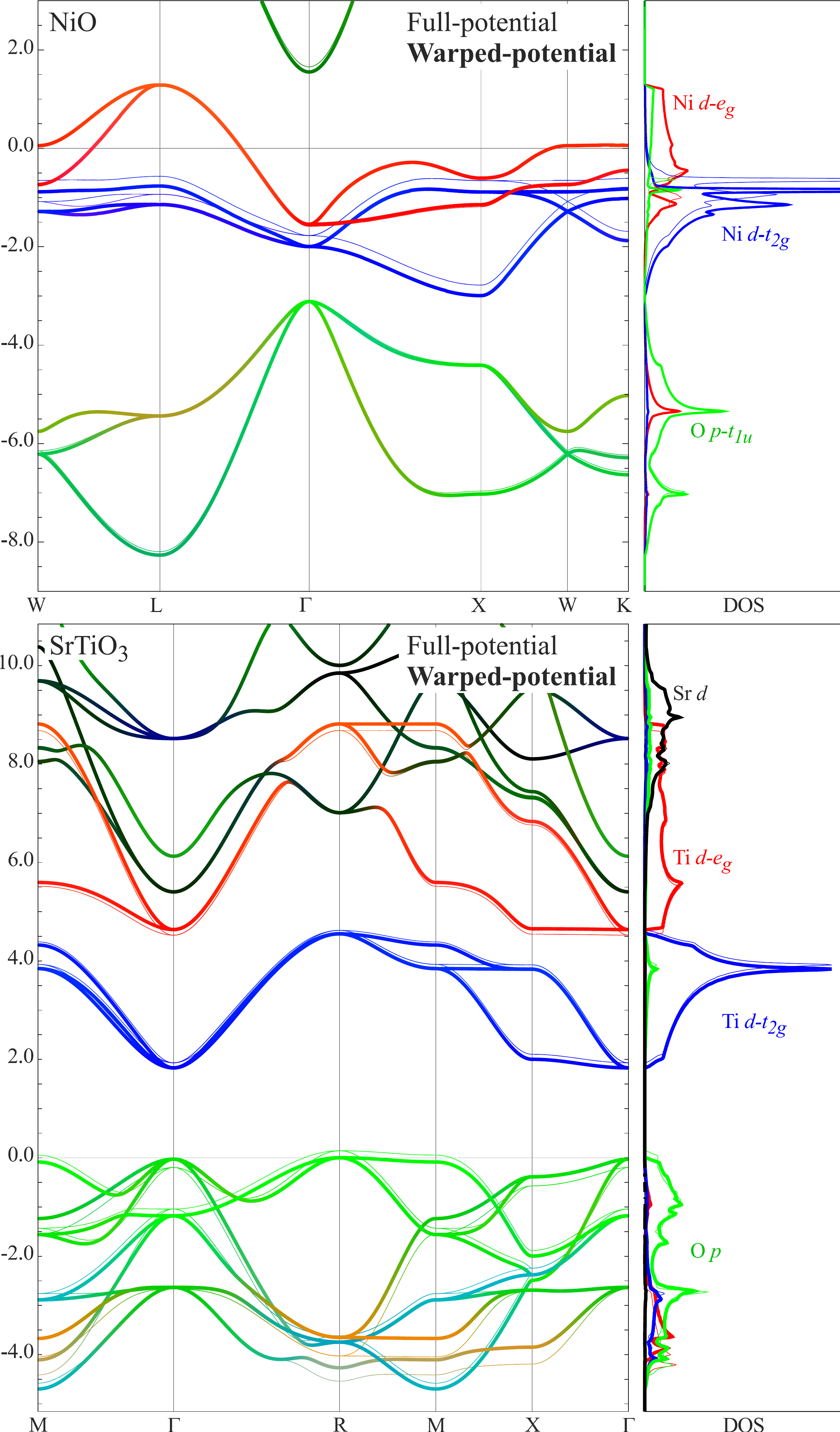}
\caption{(color online) Comparison of the LAPW band-structures and densities
of states for the full LDA potential (thin lines) and the warped MT
potential (thick lines). When only a single line can be seen, the two
band-structures overlap within the line-width of the plot. The colors
indicate the partial-wave characters inside the MT spheres.}
\label{NiOSrTiO3LDAFullWarped}
\end{figure}

DFT in the LDA already contains a large part of the local Coulomb
interactions. These interactions are included once more in the MLFT
calculations. For MLFT based on the LDA potential and Wannier orbitals one
should therefore take care not to double count such interactions. We
differentiate rigidly between the monopole and the multipole parts of all
Coulomb interactions. This idea is based on the experimental observation
that the monopole part of the Coulomb interaction ($U$) is largely screened,
from $\sim 25$ eV to $\sim 7.3$ eV in NiO for example. The multiplet
splitting, determined by the multipole part of the Coulomb interaction is,
however, only slightly reduced from the splitting one expects based on
atomic values. This has for example been observed in Auger spectroscopy for
the elemental $3d$ metals. \cite{Antonides77}

The monopole part of the Coulomb interaction ($U$) as well as the spherical
part of the on-site energy ($\Delta $) we fit to the experiment and
double-counting for the spherical part is therefore not an issue. In order
to prevent double counting of the multipole part of the Coulomb repulsion,
the LDA calculations are done with a warped LDA potential; \emph{i.e.}
within the MT sphere only the spherical part of the potential is included.
In order to check how this influences the LDA band-structure we compare in
Fig.$\,$\ref{NiOSrTiO3LDAFullWarped} the band-structures of NiO (top) and
SrTiO$_{3}$ (bottom) calculated with the full LDA potential (thick) and
warped MT potential (thin). Concentrating first on the NiO bands, we see
that both calculations agree within basically the line-width for all bands
and $\mathbf{k}$-vectors, except for the $\mathrm{t}_{2g}$ bands. Those
bands are shifted downwards in the warped-MT calculations by a momentum
independent value of about 220 meV. This effect nicely illustrates the
problem of double counting. The orbital occupation of the Ni $3d$ orbitals
within LDA is such that the $t_{2g}$ orbitals are fully occupied and the $%
e_{g}$ orbitals are half filled. The local charge density is thus cubic. The
Coulomb repulsion between two $t_{2g}$ orbitals is on average larger than
between a $t_{2g}$ and $e_{g}$ orbital. This effect is well included in the
LDA functional and related to the fact that the overlap of for example the
density of the $d_{xy}$ and $d_{xz}$ orbitals is larger than the overlap of
the density of the $d_{xy}$ and $d_{3z^{2}-1}$ orbitals. In MLFT
calculations, such interactions are included in the Slater integrals. A MLFT
calculation based on the full LDA potential would thus double count the
multipolar interaction between the fully occupied $t_{2g}$ and the half
filled $e_{g}$ shell.

One option would be to include the full potential within the LDA
calculations and then subtract the non-spherical part of the Coulomb
repulsion, as included in the LDA functional, between the Wannier functions
for which a full multiplet interaction is included in the MLFT calculations.
In that case one should carefully analyze the occupation of each Wannier
function in order to determine the potential that has to be subtracted. We
opted to not include the non-spherical interactions in the first place. This
does mean that one also neglects the non-spherical part of the non-local
Madelung potential in the self-consistent LDA calculations. In order to
correct for this, we calculate this potential from the self-consistent LDA
charge density and added it afterwards. We found that the Coulomb potential
which is double counted generally exceeds the non-spherical non-local
potential by an order of magnitude. Our choice of not including the
non-spherical interactions in the first place and treating the non-local
non-spherical interactions as a correction after self-consistency has been
reached thus provides an accurate self-consistent solution to the potential
needed in MLFT. Nevertheless, we expect that doing a full-potential
calculation and subsequently subtracting the non-spherical part of the local
Coulomb interaction according to the LDA functional will give very similar
results.

In the bottom panel of Fig.$\,$\ref{NiOSrTiO3LDAFullWarped} we compare the
band-structures of SrTiO$_{3}$ calculated for the warped and full
potentials. First of all, there is no clear shift of the $t_{2g}$ bands with
respect to the $e_{g}$ bands, presumably because Ti atom has a $\mathrm{d}%
^{0}$ configuration. But there are changes in the O $p$-derived bands. In
NiO both the Ni and the O atoms have cubic point symmetry, but in SrTiO$_{3}$
the O environment is tetragonal as was mentioned above. For SrTiO$_{3}$ one
can see a larger difference between the full and warped-potential
calculations than for NiO. The interpretation for SrTiO$_{3}$ is less
straight forward because several effects come together. Due to Ti-O
covalency there is some non-spherical, predominantly $e_{g}$-derived charge
on the Ti atom from the O $2p$ band. The local non-spherical potential due
to this charge should not be included when doing MLFT calculations. There is
however also a non-cubic potential at the O site that shifts the O bands.
This potential should be included when doing the MLFT calculations. Note
that the latter potential is not included in the self-consistent
warped-potential calculations, but we add it later, before doing the MLFT
calculations.

\section{Basis size reduction and the creation of Ligand Orbitals by blocktridiagonalization}
\label{appBasisSize}

Embedded cluster calculations, like MLFT or DMFT, contain a few correlated orbitals coupled to a large set of uncorrelated orbitals. Within such calculations the size of the Hilbert space can be reduced enormously by creating appropriate linear combinations of the uncorrelated orbitals. Within our examples these are the O $p$ orbitals, which are combined to Ligand orbitals. Without introducing ligand orbitals for a TMO$%
_{6}$ cluster, there would be $\left( 18+5\right) \times 2=46$ spin orbitals
in the one-electron basis. With a filling of $8+6\times 6=44$ electrons for
the NiO$_{6}$ $\mathrm{d}^{8}$-cluster, this results in $46!/(44!\times
2!)\sim 10^{3}$ states in the many-electron basis. For the TiO$_{6}$ $%
\mathrm{d}^{0}$-cluster representing SrTiO$_{3},$ the filling would be $%
6\times 6=$36 and thus result in $46!/(36!\times 10!)\sim 4\times 10^{9}$
states in the many-electron basis. The introduction of $L$ orbitals,
however, reduces the number of one-electron basis functions to 20, whereby
the many-electron Hilbert space reduces to $20!/(18!\times 2!)=190$ for a $%
\mathrm{d}^{8}$ and to $20!/(10!\times 10!)\sim 2\times 10^{5}$ states for a 
$\mathrm{d}^{0}$ configuration. This reduction in the number of
many-electron basis functions by factors of respectively $\frac{46!}{20!}%
\frac{\left( 20-2\right) !}{\left( 46-2\right) !}\sim 5$ and $\frac{46!}{20!}%
\frac{\left( 20-10\right) !}{\left( 46-10\right) !}\sim 22\,000$ leads to a
crucial gain of computational convenience. Either basis set \emph{can} be
used to calculate ground-state properties and spectral functions because the
matrices are sparse. But diagonalization of a matrix with dimension $4\times
10^{9}$ requires large computational resources whereas diagonalization and
evaluation of spectral functions of a sparse matrix with dimension $2\times
10^{5}$ can be done using standard libraries on modern desktop computers.
One may obtain a further reduction in the number of stored basis states by
including only those which are important for representing the actual
wave-function. (see Appendix \ref{appDiag} for details).

Ligand orbitals are normally obtained by symmetry considerations.\cite{Ballhausen62} The rotation properties of the TM $d$ orbitals should be the same as the linear combination of the O $p$ orbitals with which this orbital makes a covalent bond. These symmetry considerations can be extended to a simple mathematical procedure, valid in all point group symmetries. From DFT the one particle Hamiltonian for an extended cluster is known on a basis of the central TM $d$ orbitals and the neighbor O $p$ orbitals. Using a block Lanczos routine one can create a unitary transformation of the $p$ orbitals such that the one particle Hamiltonian has a blocktridiagonal form. The basis of the central TM $d$ orbitals is not changed. In cubic symmetry the tridiagonalization results in a transformed Hamiltonian whereby each TM $d$ orbital couples to one Ligand orbital. For lower symmetries each $d$ orbital couples to maximal 5 Ligand orbitals. The ligand orbitals can couple to an other set of ligand orbitals, \textit{ad infinitum}. Covalence in a tridiagonal representation tends to converge fast, justifying the inclusion of only a single Ligand shell in MLFT.

The introduction of Ligand orbitals is not restricted to fully occupied shells, like the O $\mathrm{p}$ shell. For example for SrTiO$_{3}$ one could include besides the O $p$ Ligand orbitals also the Sr $d$ Ligand orbitals. Care has to be taken how for such a system the Ligand orbitals are defined. If one creates a single Ligand shell for both the O $p$ and Sr $d$ orbitals by blocktridiagonalization as described in the previous paragraph one would obtain Ligand orbitals that are always partially occupied. This results in a very large many particle basis set and is unpractical. In such a case it is better to first diagonalize the non-interacting Hamiltonian describing the interactions between the Ligand orbitals. Based on the onsite energies one then creates two different Ligand shells, one for the occupied, or valence orbitals and one for the unoccupied or conduction orbitals. \cite{Gunnarsson83}

Note that a similar procedure can be used for DMFT calculations using a Lanczos impurity solver. Doing so enhances the calculation speed and allows one to increase the number of bath sites (number of discretization sites used to represent the Anderson impurity model used in DMFT) leading to much more continues spectral functions.

\section{Exact diagonalization and Lanczos algorithm}
\label{appDiag}

The MLFT ground-state and spectral calculations are done using a Lanczos
algorithm.\cite{Dagotto94,Weisse06} The calculations start with a random
vector ($\psi _{0}$) in the basis of the $d^{n}$ and $d^{n+1}\underline{L}$
configurations, whereby $n$ is the number of $\mathrm{d}$ electrons (0 for
Ti, 5 for Mn and 8 for Ni) and $\underline{L}$ represents a single hole in
the Ligand shell. Although this starting point is slightly worse than the
DFT single Slater determinant ground-state, which prescribes a specific
mixture of $d$ and $L$ states, it does contain the correct symmetry states.
Thereby convergence is so fast that the starting point really does not
matter that much. Given a negative definite Hamiltonian, the wave-function $%
\psi _{1}=H\psi_{0}$ has a larger overlap with the ground-state wave
function than the wave-function $\psi _{0}$. By repeatedly acting with the
Hamiltonian on the random starting function and normalizing the
wave-function in-between ($\psi _{n+1}=H\psi _{n}/|H\psi _{n}|$), one
converges to the ground-state. This procedure can be speeded up considerably
by creating a tri-diagonal matrix of the Hamiltonian in the basis of $\psi
_{n}$, with the additional constrained that $\psi _{n+1}$ is orthogonal to $%
\psi _{n}$. The tri-diagonal matrix in the so called Krylov basis can be
diagonalized with the use of dense matrix methods. Having found the
ground-state within the basis of the $d^{n}$ plus $d^{n+1}\underline{L}$
configurations, we remove the basis functions not needed to represent the
ground-state wave function from the basis and extend the basis set by acting
with the Hamiltonian on the wave-function. This creates basis states
belonging to the $d^{n+2}\underline{L}^{2}$ configuration. Within this new
basis set the ground-state is found and the procedure of truncating and
extending the basis set is repeated. The whole process is repeated until
convergence is reached. Excited states are calculated by repeatedly
orthogonalizing the wave-function to the eigenstates already found. The
algorithm as described here allows one to always keep relatively small basis
sets.

Spectral functions are calculated by acting with the transition operator on
the ground-state wave-function. The resulting function is then used as a
starting vector for the creation of a tri-diagonal matrix in a Krylov basis.
The spectral function of a tri-diagonal matrix can be expressed in terms of
a continued fraction.


\begin{thebibliography}{999}
\bibitem{NotLMTO} We used $N=1.$ Note that even with this value, the $N$MTO
method\cite{Andersen00} is far more physical, flexible, and accurate than
the conventional linear muffin-tin orbital (LMTO) method.\cite{Andersen84}
\bibitem{kinkcancellation} The proper wave functions at the chosen energy
are those linear combinations for which the kinks cancel.
\end{thebibliography}
\end{document}